\documentclass[%
 aip,
 amsmath,amssymb,
 reprint,%
]{revtex4-2}

\usepackage{graphicx}
\usepackage{dcolumn}
\usepackage{bm}
\usepackage{hyperref}
\usepackage[utf8]{inputenc}
\usepackage[T1]{fontenc}
\usepackage{mathptmx}
\usepackage{etoolbox}

\usepackage{subfigure}
\usepackage{url}

\usepackage{color}
\usepackage{listings}

\hypersetup{hypertex=true,colorlinks=true,linkcolor=blue,anchorcolor=blue,citecolor=blue,urlcolor=blue}

\newcommand{\abs}[1]{\left\vert#1\right\vert}

\makeatletter
\def\@email#1#2{%
 \endgroup
 \patchcmd{\titleblock@produce}
  {\frontmatter@RRAPformat}
  {\frontmatter@RRAPformat{\produce@RRAP{*#1\href{mailto:#2}{#2}}}\frontmatter@RRAPformat}
  {}{}
}%
\makeatother
\begin{document}

\preprint{AIP/123-QED}

\title{FP3D: A code for calculating 3D magnetic field and particle motion}
\author{P. Y. Jiang}
\affiliation{
Institute for Fusion Theory and Simulation and School of Physics, Zhejiang University, Hangzhou 310027, China
}
\author{Z. C. Feng}%
\affiliation{
Renewable and Sustainable Energy Institute, Department of Physics, University of Colorado, Boulder, CO 80309, USA
}%

\author{G. D. Yu}
\affiliation{%
Department of Plasma Physics and Fusion Engineering, School of Nuclear Science and Technology, University of Science and Technology of China, Hefei 230026, China.
}%

\author{G. Y. Fu}
\email{gyfu@zju.edu.cn}
\affiliation{
Institute for Fusion Theory and Simulation and School of Physics, Zhejiang University, Hangzhou 310027, China
}

\date{\today}

\begin{abstract}
An efficient numerical code FP3D has been developed to calculate particle orbits and evaluate particle confinement in 3D magnetic fields including stellarators and tokamaks with 3D fields. The magnetic field is either calculated from coils directly or obtained from equilibrium codes. FP3D has been verified with the 3D equilibrium code VMEC (S. P. Hirshman, Phys. Fluids 26, 3553 (1983)) for magnetic field calculation and with the drift-kinetic code SFINCS (M. Landreman, Physics of Plasmas 21 (4) (2014)) for neoclassical transport. The code has been applied successfully to the NCSX stellarator (B. Nelson, Fusion Engineering and design 66 (2003)) for the calculation of neoclassical transport coefficient with the 3D magnetic field obtained directly from coils. FP3D is also used to calculate ripple losses in the tokamak EAST (W. Yuanxi, Plasma Science Technology 8 (3) (2006)).
\end{abstract}

\maketitle


\section{Introduction}

Neoclassical transport is one of most important issues in stellarators. The transport coefficient scales very strongly with plasma temperature and this is not acceptable for confinement in high temperature regime if it is not reduced to a minimal level, especially for future stellarator reactors. Therefore, the neoclassical transport coefficient is a key target in stellarator design and optimization. Another important stellarator optimization target is energetic particle confinement which is very sensitive to 3D helical ripple. Typically, test particle codes \cite{McMillan_2014}\cite{HIRVIJOKI20141310}\cite{Tani_2012}\cite{Kramer_2013}\cite{wang2021ptc} or drift-kinetic codes\cite{BEIDLER1987220}\cite{wakasa2001monte}\cite{tribaldos2001monte}\cite{isaev2006venus+}\cite{kernbichler2008recent} are used for simulation of particle orbits and evaluation of particle confinement needed in stellarator optimization.

In this paper we report the development of an efficient code for Field line following and test Particle calculation in 3D magnetic fields (FP3D). The code FP3D can be used to calculate particle orbits and neoclassical confinement in stellarators as well as tokamaks with 3D fields. In particular, the code can be used to calculate magnetic field directly from 3D coils of stellarators and calculate field lines and flux surfaces. This code development is motivated by our recent work of direct stellarator optimization from 3D coils \cite{yu2021neoclassically}\cite{yu_feng_jiang_fu_2022} where a test particle code is needed for evaluation of neoclassical confinement of both thermal plasmas and energetic particles. In such direct stellarator optimization, the magnetic field needs to be calculated directly from coils.

FP3D is a template metaprogramming C++ code focusing on efficient calculation of test particle orbits and particle transport in 3D magnetic fields. The 3D magnetic field can either be calculated directly from coils or obtained from equilibrium codes such as the three-dimensional Variational Moments Equilibrium Code VMEC\cite{hirshman1983steepest}.  The field line following method is used to construct vacuum equilibrium from coils directly. It uses the guiding center equation for particle orbits and use Monte-Carlo method for particle collisions. FP3D is a versatile code with template metaprogramming and can be used in arbitrary coordinates. The Compile-time Symbolic Solver (CSS), as will be described shortly, is used to transform vector equations into component form in general curvilinear coordinates. Thus FP3D can support arbitrary equations.

This article is organized as follows:
Sec. 2 describes methods of magnetic field calculation, Sec. 3 describes calculation of magnetic flux surfaces, rotational transform and flux coordinates of stellarators as well as related benchmarks. Sec. 4 describes the equations of motion and numerical methods used in this work as well as several benchmarks;
Sec. 5 describes the structure of FP3D code; Sec. 6 describes numerical results of particle orbits, neoclassical transport coefficient in stellarators as well as ripple losses in tokamaks. Finally, summary and conclusion are given in Sec. 7.

\section{Magnetic field calculation}
\subsection{Direct calculation of magnetic field from coils}
Here magnetic field is calculated directly from coils using Biot-Savart Law given below
\begin{equation}\label{1.1}
  \vec{B}(\vec{r}) = \frac{\mu_0}{4\pi} \int \frac{Id\vec{l}\times(\vec{r}-\vec{r}_{coil})}{\abs{\vec{r}-\vec{r}_{coil}}^{3}} \approx  \frac{\mu_0 I}{4\pi} \sum_{i} \frac{(\vec{r}_{i+1}-\vec{r}_i)\times(\vec{r}-\vec{r}_{i+1/2})}{\abs{\vec{r}-\vec{r}_{i+1/2}}^{3}}
\end{equation}
where integration is done section by section along each coil's path and $\vec{r}_{i+1/2}=(\vec{r}_{i}+\vec{r}_{i+1})/2$. This method can calculate the magnetic field at any position with high accuracy if the length of each section is short enough. The typical number of grid points of a single 3D coil is 500 which corresponds to a relative error of $10^{-5}$. For special cases such as Columbia Non-Neutral Torus (CNT)\cite{pedersen2004columbia} which has four circular coils, the elliptic function is employed to calculate the magnetic field. This special method is much faster than coil integration.

\subsection{Interpolation method for calculating magnetic field} \label{sec2.2}
In advanced stellarators such as Wendelstein 7-X (W7X)\cite{hartfuss1997diagnostic}, the shapes of coils are complex and the number of coils is large. This means that the number of grid points needed for coil integration is large for sufficient accuracy and the computational cost is high. Therefore grid interpolation is used for both accuracy and efficiency. Furthermore, interpolation is also necessary to evaluate field from field grid data generated by equilibrium codes.

FP3D uses uniform grids in cylindrical coordinates. The 3-dimension B-spline interpolation \cite{bspline-C++} is employed whose order can be selected from 3 to 8. The interpolation produces function values and partial derivatives of functions at any point in space needed for calculating the gradient and divergence terms in the particle equations of motion.

\begin{figure}[htbp]
  \centering
  \subfigure{
    \includegraphics[width=0.3\textwidth]{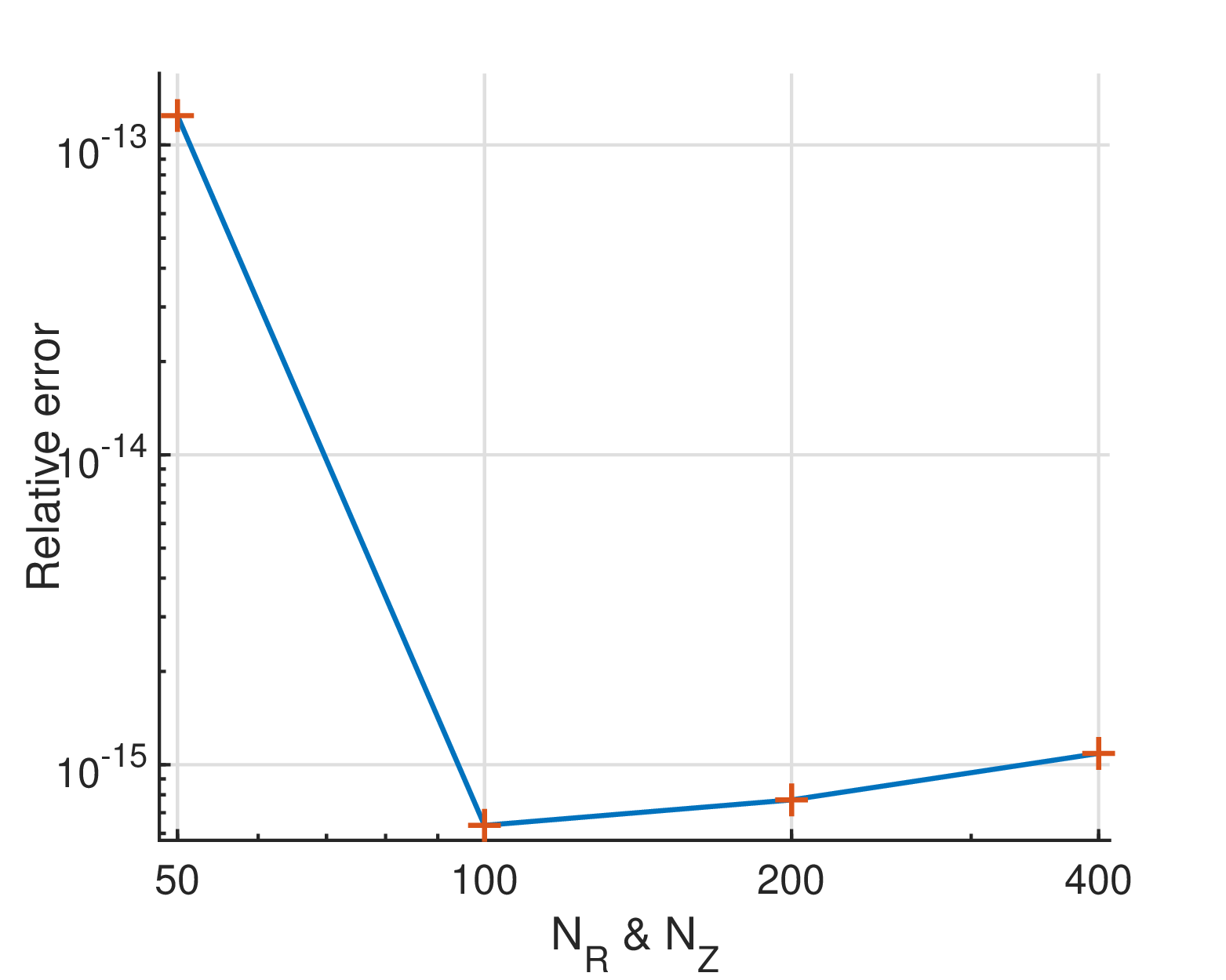}}
  \subfigure{
    \includegraphics[width=0.3\textwidth]{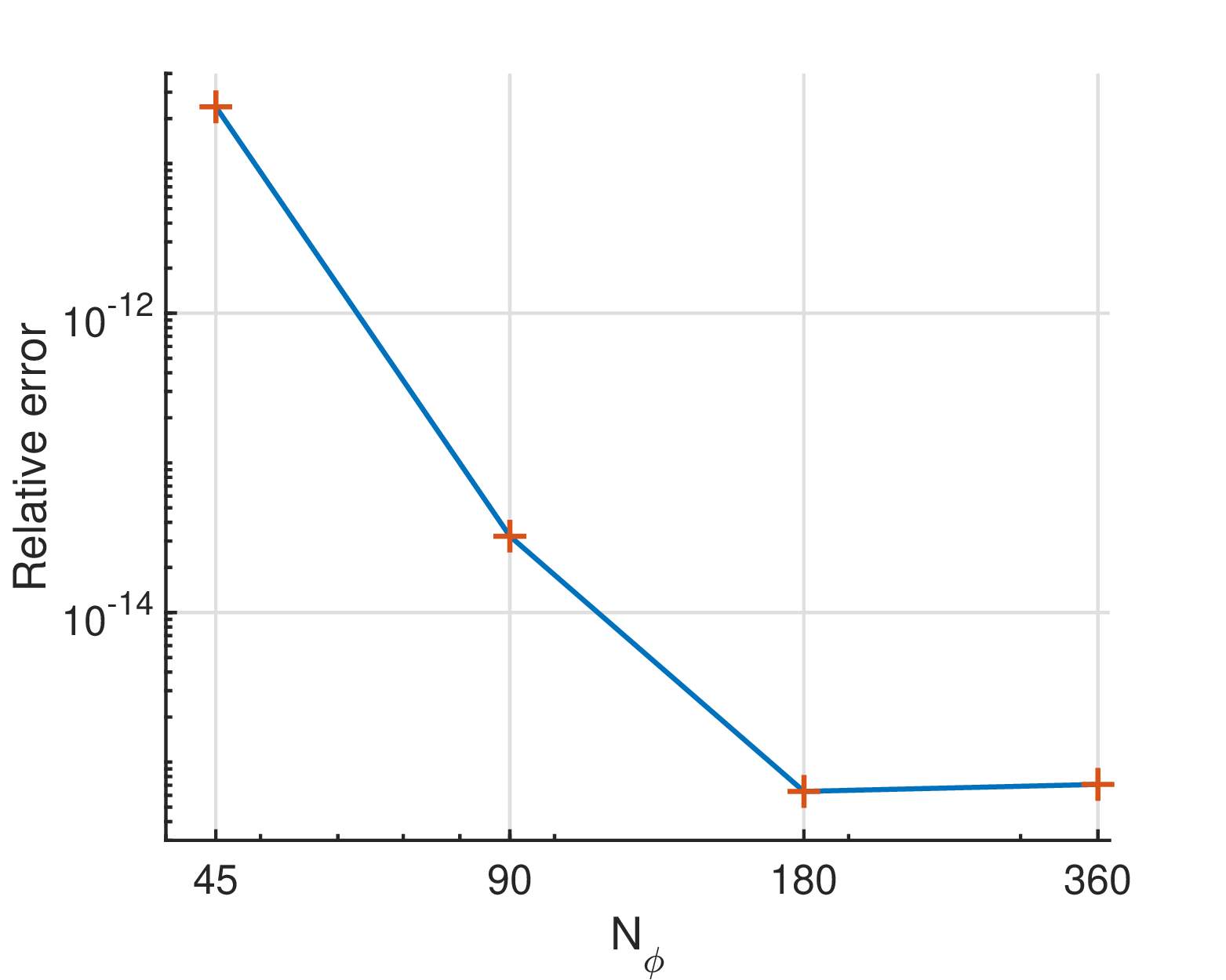}}
  \subfigure{
    \includegraphics[width=0.3\textwidth]{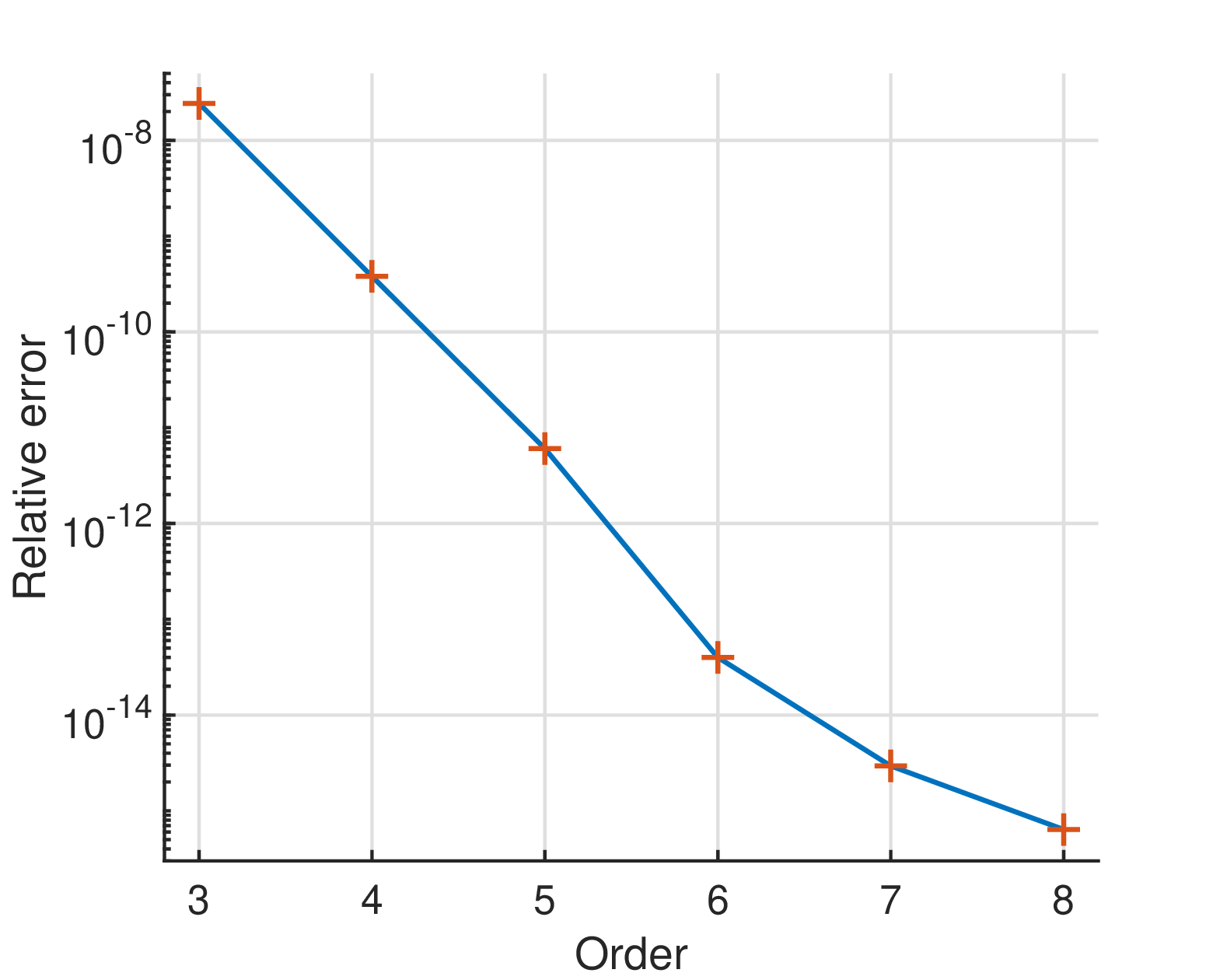}}
  \caption{The averaged relative error as function of grid number $N_R$, $N_Z$, $N_\phi$ and the order of interpolation for magnetic field of CNT.}
  \label{fig:Griderror}
\end{figure}

\textbf{Benchmark}
We compare the interpolated off-grid values of magnetic field with those calculated directly from coils of CNT. The average relative errors are shown in Fig \ref{fig:Griderror}. For the baseline case of $(N_R,N_Z,N_{\phi})=(100,100,180)$ and 8th order where $N_i$ is the number of grid points in $i$ direction, the relative errors are less than $10^{-15}$, which is close to double precision.

\subsection{Simple model for calculating electric field}
For 3D stellarators or tokamaks with 3D fields, the difference between ion and electron transport causes charge separation and generates electric field. The electric field reduces ion transport greatly to achieve ambipolar transport. FP3D allows a static electric field. The electric field is a function of flux coordinate $(\psi,\theta,\phi)$ in general, but for simplicity, we assume that electric potential $\Phi$ only depends on $\psi$, $\Phi = \Phi(\psi)$. Thus the electric field can be expressed as
\begin{equation}\label{1.2}
  \vec{E}=\Phi'(\psi)\nabla\psi
\end{equation}
where $\Phi'(\psi)$ is expanded in polynomials.

\subsection{Interface with equilibrium codes}
We have developed an interface with the output of the equilibrium code VMEC so that the magnetic field from VMEC can be used. VMEC uses flux coordinates $(\psi,\theta,\phi)$ where $\psi$ is toroidal flux, $\theta$ and $\phi$ are poloidal angel and toroidal angle respectively. Mapping them to cylindrical grids would lead to large errors because of the mismatch of boundary shapes. FP3D supports any coordinates including curvilinear coordinates. In the above cases of magnetic field from coils, the magnetic field is saved in cylindrical grids but field line following and particle calculation are done in Cartesian coordinates. Here, FP3D reads equilibrium field grids directly and all the computation is done in flux coordinates. An external MATLAB program has been implemented for calculating equilibrium quantities such as metric tensor needed by FP3D in VMEC coordinates.

\section{Calculation of Magnetic surfaces, rotational transform and flux coordinates of stellarators}
\subsection{Magnetic surfaces and rotational transform}
Magnetic surfaces are important for plasma confinement, and can be calculated by tracing field lines. The trajectory of a magnetic field line $\vec{r}(l)$ satisfies the following equation
\begin{equation}
  \frac{d\vec{r}}{dl} =\frac{\vec{B}}{\abs{\vec{B}}} \label{2.1} \\
\end{equation}
or
\begin{equation}
  \frac{d\vec{r}}{d\phi}  =\frac{\vec{B}}{\vec{B}\cdot\nabla\phi} \label{2.2}
\end{equation}
where $l$ is the length along the field line.
By solving ordinary differential equation \ref{2.1}, we can obtain field lines for any magnetic field. This method is useful to judge whether magnetic surfaces exist in stellarator optimization. On the other hand, Eq. \ref{2.2} is convenient for constructing Poincare surfaces at a fixed value of $\phi$ as long as field lines form magnetic surfaces.

\textbf{Benchmark}
It is well known that the magnetic field lines of a single circular coil are closed. For a coil of radius a=1m , we start from a location in the coil plane 0.8 meter away from the coil center and trace $10^5$ laps around the coil, the last point where field line intersects the coil plane is $2\times10^{-6}m$ away from the starting point. This indicates that the field line following is very accurate.

\begin{figure}[ht]
\centering
\includegraphics[width=0.5\textwidth]{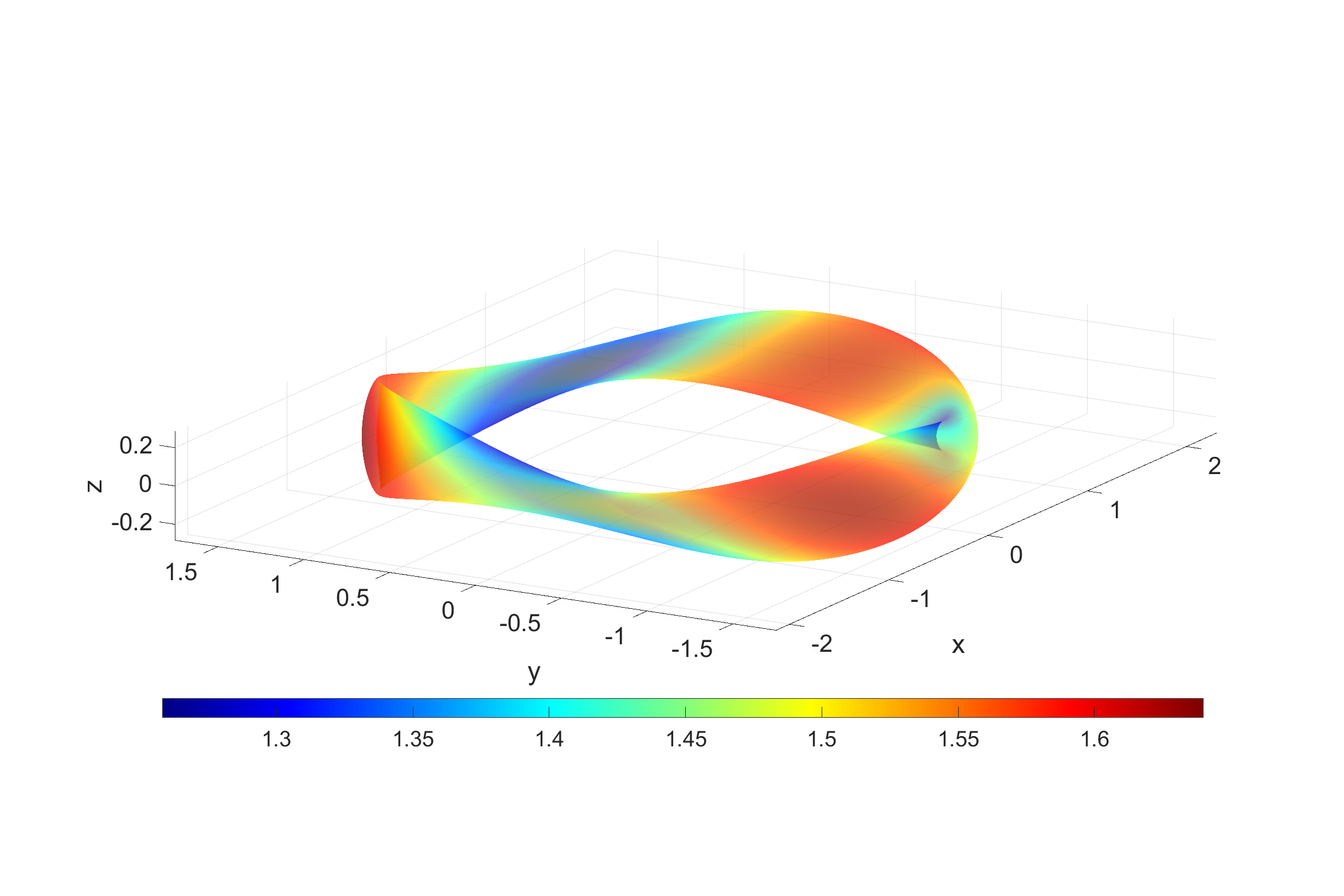}
\caption{The calculated magnetic surface with initial tracing point $(x,y,z)=(1.64,0,0)$. The color-bar represents the strength of local magnetic field.}
\label{fig:MGF}
\end{figure}

\begin{figure}[ht]
\centering
\includegraphics[width=0.5\textwidth]{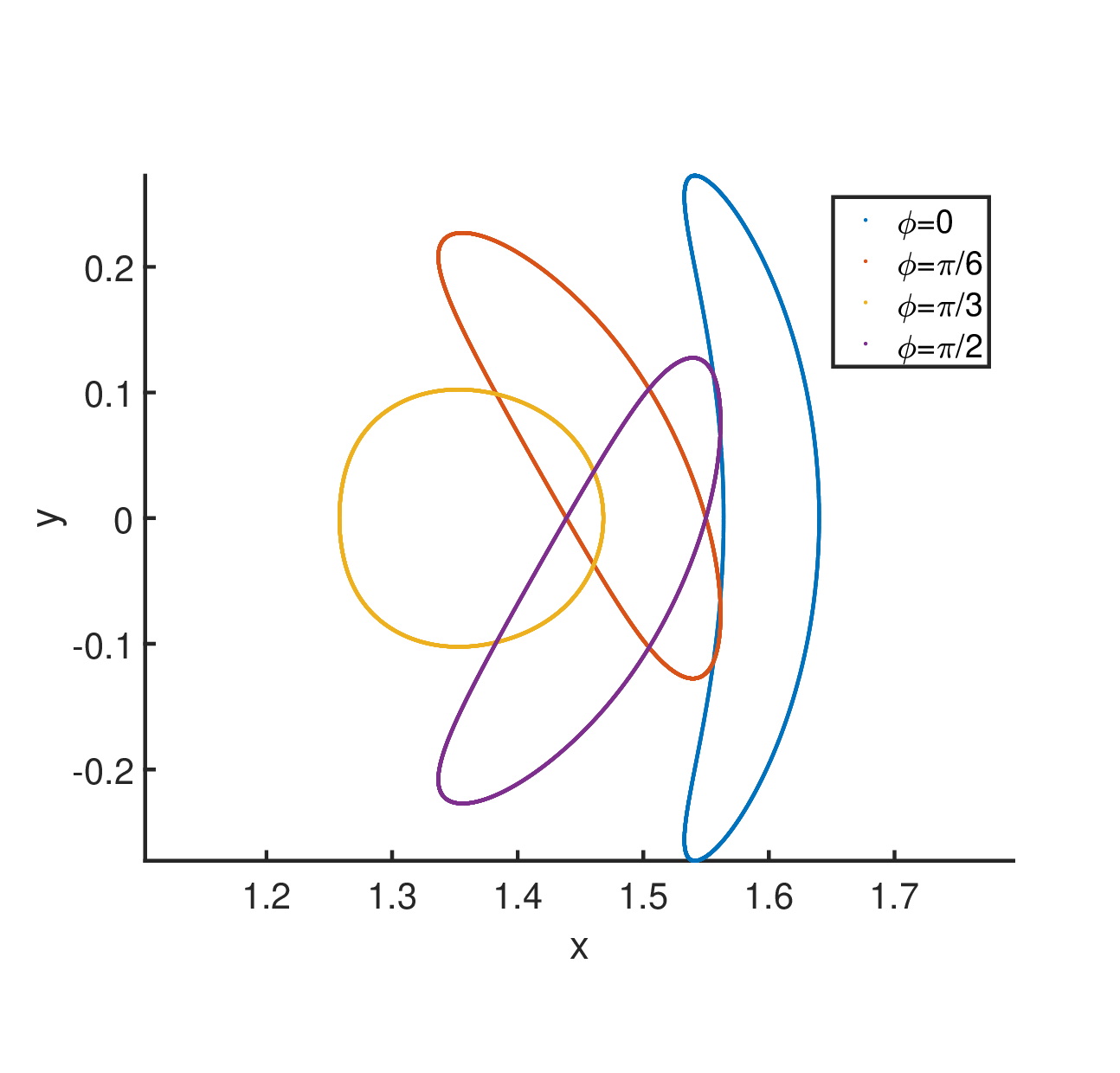}
\caption{The Poincare sections of NCSX stellarator. Key parameters in the calculations are $r_0=(1.64,0,0)$, step length $d\phi = 2\pi/360$, and 1000 tracing laps.}
\label{fig:poincare}
\end{figure}

For stellarators, using equal $\phi$ method, the magnetic surfaces made of field lines are closed 3D surfaces. The Poincare section(at $\phi=const$) forms a closed curve. Figure \ref{fig:MGF} and \ref{fig:poincare} show a calculated magnetic surface and Poincare sections of National Compact Stellarator Experiment (NCSX)\cite{nelson2003design} with 21 coils (7 coils for each period). 500 grid points are used in the calculation of magnetic field from each coil using Eq (\ref{1.1}).

Magnetic surfaces consist of a series of nested 3-dimensional surfaces, the center one reduces to a single line called magnetic axis. If we trace a field line from $(R,Z,\phi_0)$ to $(R',Z',\phi_0+2\pi)$, the distance between them is $\Delta d$ ($\Delta d=0$ for the magnetic axis). We use a nonlinear optimization algorithm HYBRD\cite{10.1007/BFb0067700} to find the minimum of the function $\Delta d(R,Z)$ which corresponds to the magnetic axis. The merit of the algorithm is fast convergence except for the special case of $\iota=1$.

The rotational transform, labeled $\iota$, is the ratio of the times a magnetic field line travels poloidally (the "short way") to the times the magnetic field travels toroidally (the ``long way")\cite{wiki2.org}. We define poloidal angle  $\theta=\tan(Z-Z_a)/(R-R_a)$ where $(R,Z)$ and $(R_a,Z_a)$ are the cylindrical coordinates for the target magnetic surface and magnetic axis respectively. We trace one field line for the magnetic surface and another one for the magnetic axis using following equations simultaneously \cite{todoroki2003calculating}.
$$
\left\{
             \begin{aligned}
             \frac{dR}{d\phi} & = \frac{\vec{B}(R,\phi,Z)\cdot\nabla R}{\vec{B}(R,\phi,Z)\cdot\nabla \phi}  \\
             \frac{dZ}{d\phi} & = \frac{\vec{B}(R,\phi,Z)\cdot\nabla Z}{\vec{B}(R,\phi,Z)\cdot\nabla \phi}  \\
             \frac{dR_a}{d\phi} & = \frac{\vec{B}(R_a,\phi,Z_a)\cdot\nabla R}{\vec{B}(R_a,\phi,Z_a)\cdot\nabla \phi}  \\
             \frac{dZ_a}{d\phi} & = \frac{\vec{B}(R_a,\phi,Z_a)\cdot\nabla Z}{\vec{B}(R_a,\phi,Z_a)\cdot\nabla \phi}  \\
             \frac{d\theta}{d\phi} & = \left( \left(\frac{dZ}{d\phi}-\frac{dZ_a}{d\phi}\right)(R-R_a)-\left(\frac{dR}{d\phi}-  \frac{dR_a}{d\phi}\right)(Z-Z_a) \right)/\rho^2
             \end{aligned}
\right.
$$
where $\rho^2=(R-R_a)^2+(Z-Z_a)^2$, the derivatives in the fifth equation can be derived from the first four equations. The rotational transform can then be calculated as $\iota=\rm{lim}_{\Delta\phi\to\infty}\Delta\theta/\Delta\phi$ where $\Delta\theta$ and $\Delta\phi$ is the accumulative change of $\theta$ and $\phi$ respectively. The $\iota=\Delta\theta/\Delta\phi$ converges very well after tracing 50 toroidal turns.

\begin{figure}[ht]
\centering
\includegraphics[width=0.5\textwidth]{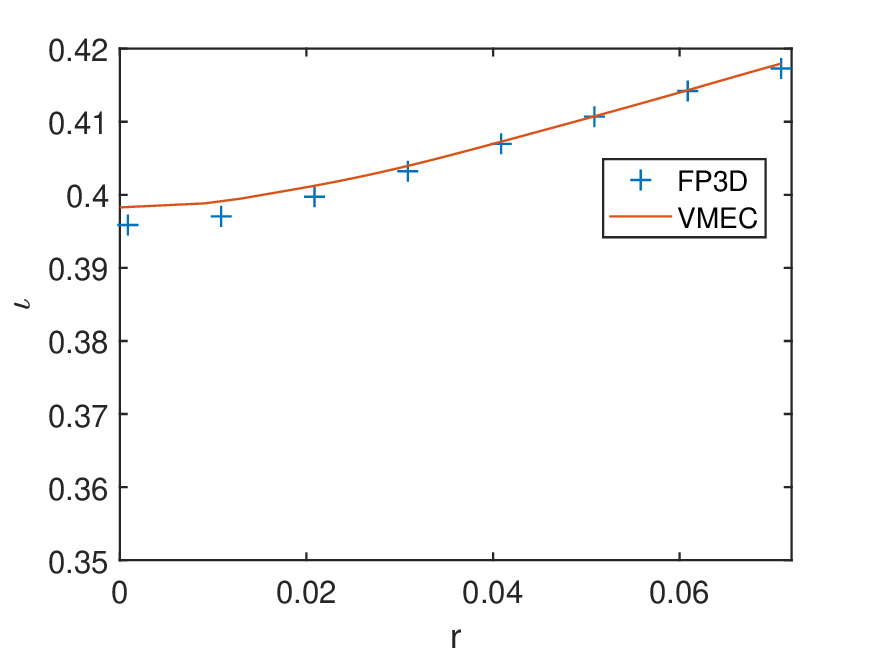}
\caption{The converged $\iota$ versus minor radius. The red line is calculated by VMEC. Blue crosses are obtained by FP3D.}
\label{fig:VMECiota}
\end{figure}

\textbf{Benchmark}
Figure \ref{fig:VMECiota} compares the FP3D's results of rotational transform (blue crosses) with those of VMEC code (red line) for NCSX. The input parameters of VMEC are the Fourier coefficients of the outmost magnetic surface with zero plasma pressure. We see that FP3D's results agree well with VMEC's.

\subsection{Magnetic flux coordinates}
Magnetic coordinates are important for calculating transport coefficient. It is defined by ($\psi,\theta,\phi$) where $\psi$ is toroidal flux, $\theta$ and $\phi$ are poloidal angel and toroidal angle respectively.

To calculate $\psi$ for the outermost magnetic surface, we locate the magnetic axis firstly. Then we divide the straight line between the starting point of outmost magnetic surface and magnetic axis equally to generate radial grids for magnetic surfaces. Second, for each of grid points, we trace field line by equal-$\phi$ method to find corresponding magnetic surface. The number of integration steps in tracing is sufficiently large to ensure the points cover the whole magnetic surface. In doing so care must be taken to avoid rational magnetic surfaces. Finally, for every magnetic surface, we rearrange the intersection points in the starting plane (at $\phi=0$) so that the points are sequential poloidally. The points in other $\phi$ plane will rearrange automatically because the magnetic lines can not intersect. We now can calculate the magnetic flux of each surface by using following equation:
\begin{equation}\label{2.3}
  \psi=\iint \vec{B}\cdot d\vec{S}=\oint \vec{A}\cdot d\vec{l}
\end{equation}
where $\vec{A}$ is the vector potential of magnetic field and $\vec{A}$ can be calculated in similar way as in Eq(\ref{1.1}). In the field line following, we get many data points $\psi_i(R_{ijk},Z_{ijk},\phi_k)$ where i is the index of magnetic surface, $j$ is the poloidal index and $k$ is the toroidal index. For given $i$ and $k$,
$\theta_j$ is defined using poloidal arc length and calculated by
\begin{equation}
  \theta_j \approx 2\pi\frac{\sum_{0}^{j} \sqrt{(R_{j+1}-R_{j})^2+(Z_{j+1}-Z_j)^2}}{\sum_{0}^{N} \sqrt{(R_{j+1}-R_{j})^2+(Z_{j+1}-Z_j)^2}}
\end{equation}
Now we can calculate $(\psi,\theta,\phi)$ at any position inside the outermost magnetic surface by scatter interpolation. However, this method can not always ensure accuracy because scatter interpolation is less accurate than uniform interpolation. A better way is to set up coordinate grid by high order scatter interpolation method firstly and then interpolate to any position by uniform grid interpolation method(same as in section \ref{sec2.2}).

The cubic 'griddata' method of MATLAB\cite{luis2007mirone} is used for accuracy to generate the uniform cylindrical grid values for $\psi_i$. This scatter interpolation needs to be done only once for each equilibrium.

\begin{figure}[ht]
\centering
\includegraphics[width=0.5\textwidth]{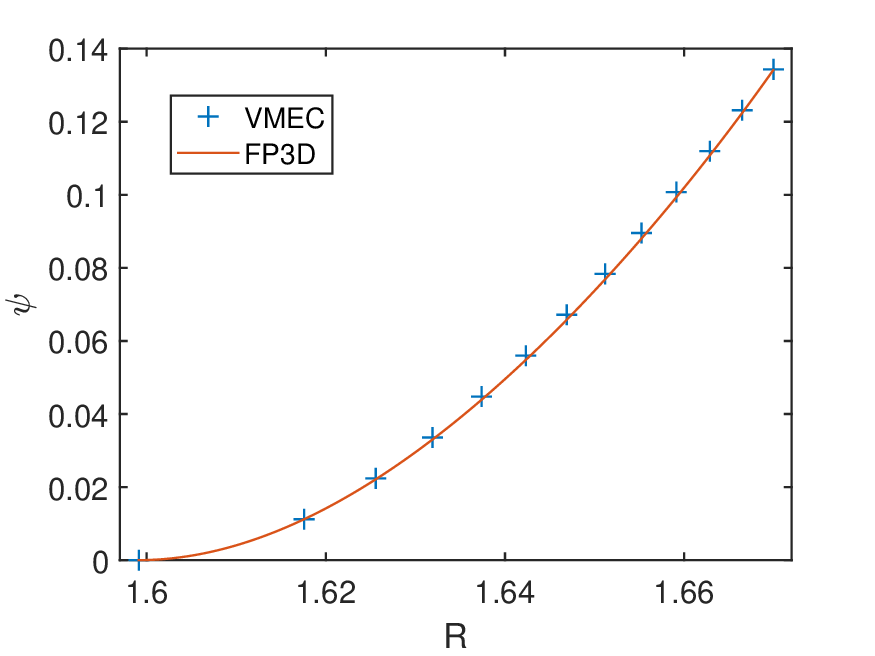}
\caption{Toroidal flux $\psi$ as a function of radius. The red line is obtained from FP3D, the blue crosses are VMEC results.}
\label{fig:VMECpsi}
\end{figure}

\textbf{Benchmark}
Since $\psi$ is constant on each magnetic surface, we trace field line and calculate $\psi$ value at every step via interpolation. The average 'relative' error of $\psi/\psi_{edge}$ is $5\times10^{-5}$ for NCSX stellarator. This error mainly comes from scatter interpolation rather than B-spline interpolation. We also compare our calculated $\psi$ with that of VMEC. The results are shown in Fig.\ref{fig:VMECpsi}. Our results agree very well with VMEC's.

\label{sec3.2}
An analytical expression of magnetic surface is very useful as a bridge between FP3D and other codes. For a three-dimensional surface which is a function of two independent angles, the Fast Fourier Transform (FFT) is used. By following the above method we get ordered poloidal points $\vec{r}(\theta',\phi)$ in a given magnetic surface.  Here the poloidal angle defined as $2\pi\sum_{j=1}^{i}\Delta l_j/l_{total}$ with $\Delta l$ being the distance between two adjacent points and $l_{total}$ being the total poloidal arc length at a given $\phi$. The magnetic surface can then be expressed as
\begin{flalign}
  R&=\sum_{m,n}R_{m,n}\rm{cos}\left(m\theta'-n\phi\right)\label{2.4}  \\
  Z&=\sum_{m,n}Z_{m,n}\rm{sin}\left(m\theta'-n\phi\right)\label{2.5}
\end{flalign}
where $R_{m,n},Z_{m,n}$ are the Fourier coefficients.


\section{Simulation of test particle motion in 3D magnetic fields}
\subsection{Particle equation of motion and time advance} \label{sec4.1} 
In magnetic confinement devices such as tokamaks and stellarators, the characteristic cyclotron radius of thermal ions and energetic ions is typically much smaller than characteristic scale of magnetic field. Thus the drift kinetic equation\cite{littlejohn1981hamiltonian} is a good approximation for particle motion and is used in FP3D. The equations are given below,
\begin{flalign}
    \frac{d\mathbf{X}}{dt} & = \frac{U_\parallel}{B^*_\parallel}\mathbf{B}^*+
    \frac{1}{qBB^*_\parallel}(\mu\mathbf{B}\times\nabla B + q\mathbf{E}\times\mathbf{B})\label{3.1} \\
    \frac{dU_\parallel}{dt} &= \frac{1}{mB^*_\parallel}\mathbf{B}^*\cdot(q\mathbf{E}-\mu\nabla B)\label{3.2}
\end{flalign}
where
\begin{flalign}
\mathbf{B}^* &=\mathbf{B}+\frac{mU_\parallel}{q}\nabla\times\mathbf{b} \label{3.3} \\
B^*_\parallel &= \mathbf{b}\cdot\mathbf{B}^*=B+\frac{mU_\parallel}{q}\mathbf{b}\cdot\nabla\times\mathbf{b}\label{3.4}
\end{flalign}
and $\mathbf{b}=\mathbf{B}/B$, $\mu=mv^2_\perp/2B$. The particle variables in guiding center equation are $(\mathbf{X},U_\parallel)$, where $U_\parallel$ is parallel component of velocity along magnetic field direction, and the $\nabla$ operator can be calculated by numerical differentiation or grid differential interpolation. The number of independent variables is fewer than that of full orbit equations. The step size of time advance can be large because the gyro-motion is ignored. Thus it is much faster to solve the drift-kinetic equation than the full orbit equation.

All above equations are generated in the code by the Compile-time Symbolic Solver (CSS) which uses C++ template metaprogramming method. The main process of CSS is summarized as following: (1) represent all vectors in component form, (2) convert vector operations in curvilinear coordinates system to scalar operators using component expressions, (3) use binary expression trees to express scalar arithmetic and derivation operations, (4) evaluate the binary expression trees at runtime. There is no extra cost at runtime because all the symbolic operations are done at compile time, the commands executed by CPU are completely same as if one writes equations in component form. This means that equations can be modified easily. Meanwhile, CSS supports arbitrary coordinate systems and arbitrary boundary conditions. The advantage of CSS for coding is easier programming and more efficient numerical operations. The details of CSS will be reported elsewhere.


The FP3D code supports both Cartesian coordinate system and curvilinear coordinate system $(\psi,\theta,\phi)$. These equations can be written in C++ code shown below
\begin{lstlisting}
	constexpr auto curl_b_con = cross(Nabla, b_cov);
	constexpr auto E_star_cov = E_cov - mu / q * (Nabla * B_s);
	constexpr auto B_star_con = B_con + m / q * U * curl_b_con;
	constexpr auto B_starp_s = b_cov * B_star_con;
	constexpr auto dr_dt = (U * B_star_con - cross(b_cov, E_star_cov));
	constexpr auto dU_dt = q / m * B_star_con * E_star_cov;
\end{lstlisting}
where 'Nabla' is $\nabla$, 'b' is $\mathbf{b}$, 'E\underline{~}star', 'B\underline{~}star', 'B\underline{~}starp' represent $\mathbf{E}^*,\mathbf{B}^*,B^*_\parallel$ respectively, and the subscript 's', 'cov' and 'con' denote scalar, covariant vector and contravariant vector respectively. 'cross' function is cross product. These combinations of different vector forms is to avoid metric components in expanded expressions which accelerates computing significantly. The result of CSS can be evaluated in following code
\begin{lstlisting}
	auto B_starp_v = symbolic_evaluate(B_starp_s, inpproxy(coor));
	auto dr_dt_v = symbolic_evaluate(dr_dt, inpproxy(coor));
	auto dU_dt_v = symbolic_evaluate(dU_dt, inpproxy(coor));
	dy_dt[0] = dr_dt_v[0] / B_starp_v;
	dy_dt[1] = dr_dt_v[1] / B_starp_v;
	dy_dt[2] = dr_dt_v[2] / B_starp_v;
	dy_dt[3] = dU_dt_v / B_starp_v;
\end{lstlisting}
where `coor' is the current particle coordinates, and `inpproxy(coor)' helps function `symbolic\underline{~}evaluate' to evaluate symbolic expression values.

We use Runge-Kutta scheme(the order can be chosen from 4th to 8th)\cite{dormand1980family} or LSODE(Livermore Solver for Ordinary Differential Equations)\cite{radhakrishnan1993description} \cite{hindmarsh2005lsoda} library in time advance. FP3D uses Adams method (predictor-corrector) in LSODE because the equations for both field line tracing and particle motion are nonstiff. The time step size is self-adaptive so that the error tolerance can be set arbitrarily.

\begin{figure}[ht]
\centering
\includegraphics[width=0.5\textwidth]{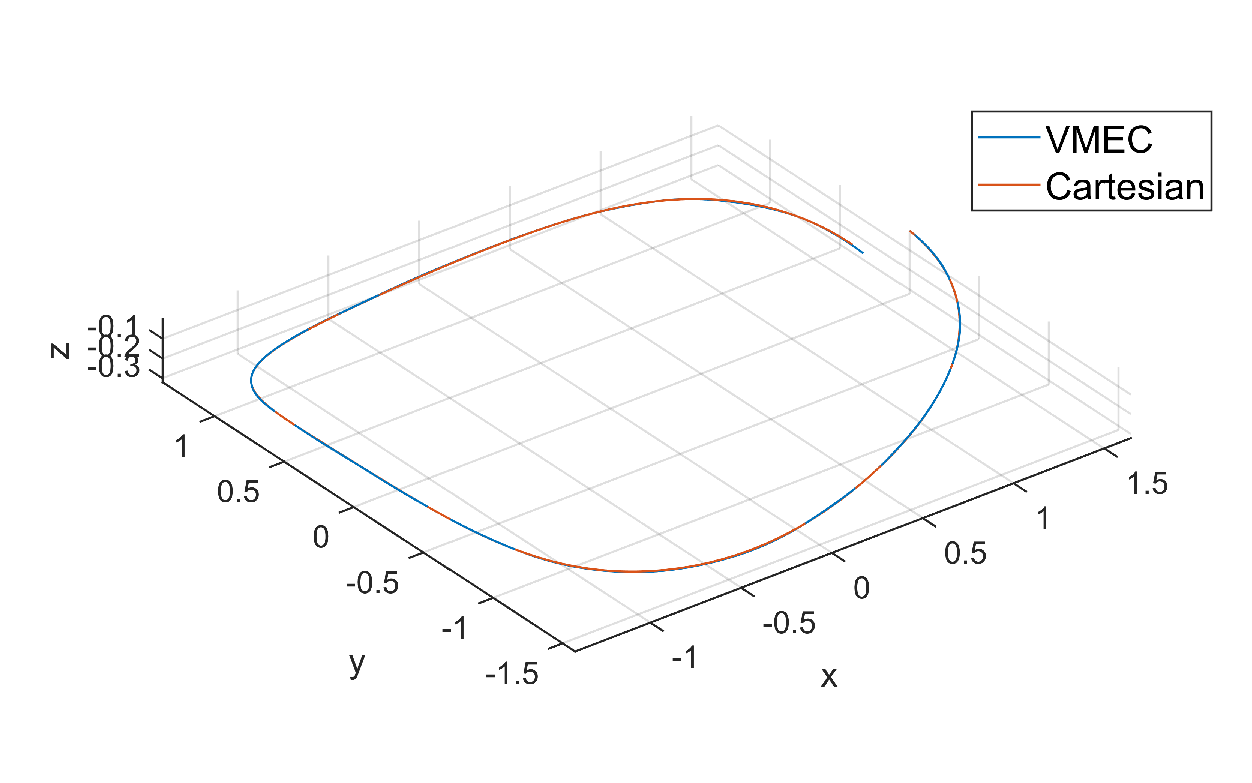}
\caption{Passing particle orbits calculated in VMEC flux coordinates (blue line) and Cartesian coordinates (red line)
respectively. The initial position is $(\psi,\theta,\phi)=(0.5,0,0)$. The particle energy is 1keV with $\mu = 0$.}
\label{fig:Orbit_compare}
\end{figure}

\textbf{Benchmark}
Here we verify the orbit calculation in magnetic flux coordinates. We use FP3D to calculate the Fourier coefficients of the outmost magnetic surface based on the 3D magnetic field obtained from NCSX coils. The coefficients are used as the input parameters for VMEC for calculating the vacuum equilibrium. Then the equilibrium output of VMEC is used to generate curvilinear grids and magnetic field in flux coordinates for FP3D. Figure \ref{fig:Orbit_compare} shows particle orbits starting from the same initial position but calculated in different coordinates with different field calculating method. The blue orbit is calculated in VMEC's flux coordinates $(\psi,\theta,\phi)$ with magnetic field from VMEC equilibrium. The red orbit is calculated in Cartesian coordinate $(x,y,z)$ with magnetic field obtained directly from coils. The results are almost the same. The small difference comes from small difference between the magnetic field from VMEC and the magnetic field calculated directly from coils. These results verify our method of orbit calculation in flux coordinates.

\subsection{Method for particle collision}
Particle collisions are simulated using the Monte Carlo method\cite{boozer1981monte}. The pitch angle is changed from $\lambda_0$ to $\lambda_n$ after a time step of $\Delta t$ with
\begin{equation}\label{3.5}
  \lambda_n=\lambda_0 ( 1-\nu_d \Delta t ) \pm \left[(1-\lambda_0^2)\nu_d \Delta t \right]^{1/2}
\end{equation}
where the scattering collision frequency $\nu_d$ is $\nu_d = 1.182\nu_B$, with $\nu_B$ being the Braginskii collision frequency
\begin{equation}\label{3.5.1}
  \nu_B=\frac{4}{3}\left(\frac{\pi}{m}\right)^{1/2}\frac{\ln\Lambda e^4n}{T^{3/2}}=4.7140\times10^{-8}\frac{n \ln\Lambda}{A^{1/2}T^{3/2}}
\end{equation}
In this formula $\ln\Lambda$ is the Coulomb logarithm, $A$ is the atomic mass of the ions, $n$ is the electron density (per $cm^3$), and $T$ is the temperature in electron volts. The convergence of algorithm requires $\nu_d \Delta t \ll 1$. The symbol $\pm$ means that the sign is to be chosen randomly, with equal probability for plus and minus.

In our code, particles' initial conditions include position vector $\vec{r}$, energy $E$, the normalized parallel velocity $U_n$ (normalized by total velocity). These parameters can be specified in input or generated from a specific distribution.
For particle initial positions, loading a uniform spatial distribution on a specific magnetic surface is an important function of the code. First, we obtain the Fourier expression of the magnetic surface $\vec{r}(\theta,\phi)$ as shown in subsection \ref{sec3.2}. Then, using Eq. \ref{2.4} and Eq. \ref{2.5}, we obtain the surface element $dS$ as
\begin{equation}\label{3.8}
  dS=\abs{\frac{\partial\vec{r}}{\partial\theta}\times\frac{\partial\vec{r}}{\partial\phi}}d\theta d\phi
\end{equation}
Finally, we use Monte Carlo method to generate a uniform distribution of $dS$ based on jacobi. For energy, the Maxwellian distribution can be generated by Monte Carlo method. And for $U_n$, an isotropic distribution is used for thermal ions or electrons.

The particle boundary is important to determine escaped particles. The code provides many ways of specifying particle boundary such as analytical boundary, cubic boundary in Cartesian coordinates, cylindrical boundary and the boundary of the outermost magnetic surface.



\section{Execution flow chart}



\begin{figure}[ht]
\centering
\includegraphics[width=0.5\textwidth]{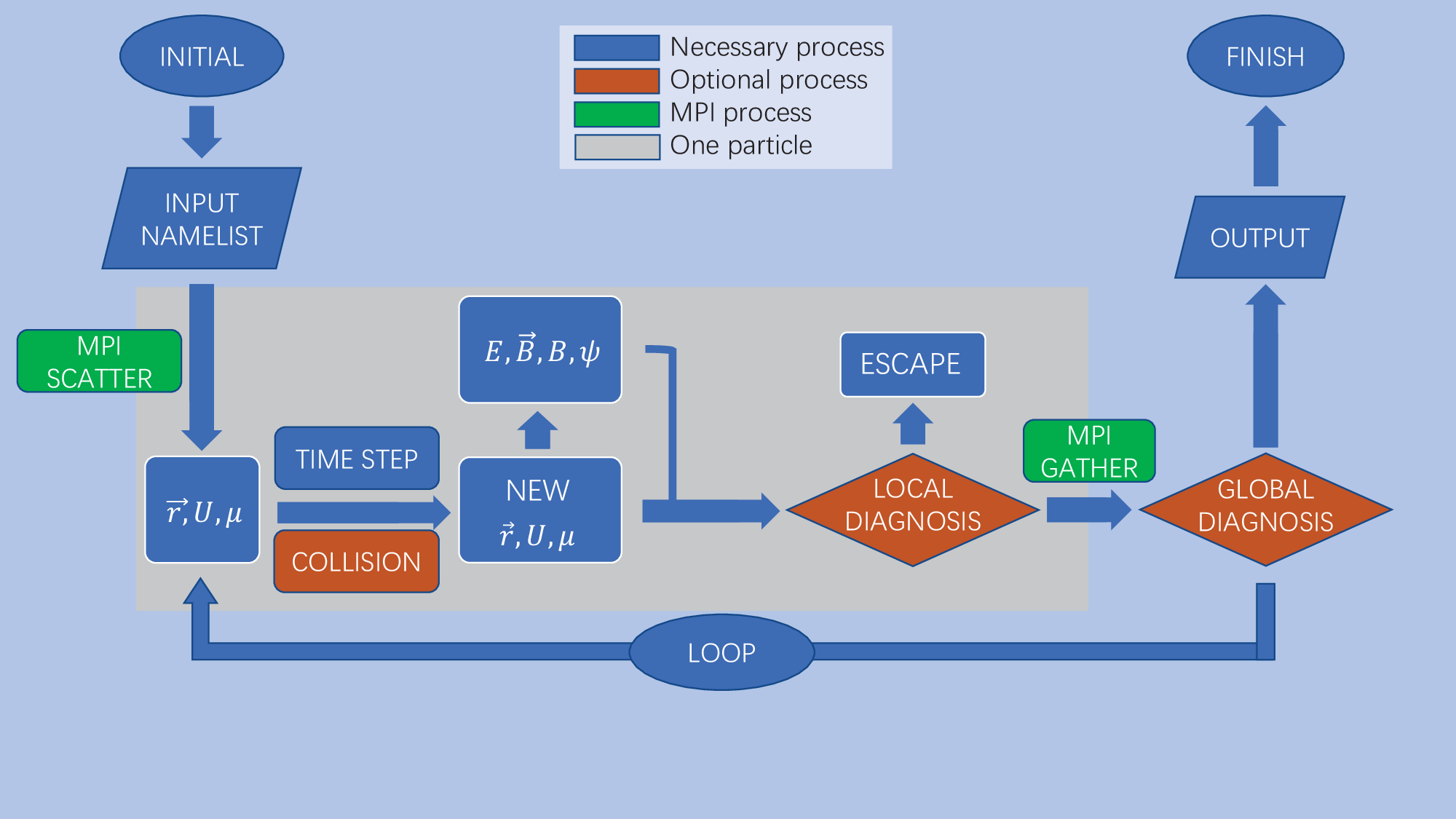}
\caption{Main modules of particle motion function. The modules in the blue blocks are necessary, and those in the origin blocks are optional. The green blocks indicate the MPI directives if it runs in a parallel environment. The processes in gray box are done for every particle.}
\label{fig:flowchart}
\end{figure}

The basic flow chart is shown in Fig.\ref{fig:flowchart}. At code initialization, the initial input files are read first to choose necessary functions. The format of input parameter file is JSON\cite{Lohmann_JSON_for_Modern_2022} because of convenience for both read and modification. the code reads analytical equilibrium data or grid information of $\vec{B},\vec{E},\psi$. Then it reads particle initial conditions or generates an appropriate distribution and transform it in $\vec{r},U,\mu$ for the ODE solver. The root process sends them to other processes if MPI enable. For each time step, the solver updates particle variables, then executes collision operation if required. The code calculates $E,\vec{B},|\vec{B}|,\psi$ at new position and save necessary values in history buffer array. Then the code runs local diagnose functions such as boundary to judge whether each particle is lost or not. When particle advance is completed, global diagnose functions will be executed to gather statistical data likes the averaged radial spread $\overline{\Delta\psi^2}$. All the diagnose and history functions are called at fixed time intervals with output of particle orbits and other information automatically.

All functions are integrated in the code which is controlled by different input parameter files. All primary modules are object-oriented with specific interfaces and can be extended or overloaded easily. For example, when calculating the diffusion coefficient $D_{\psi\psi}$, a condition can be added for $\overline{\Delta\psi^2}_{max}$ in global diagnose function to end program if $\overline{\Delta\psi^2}\geq \overline{\Delta\psi^2}_{max}$.

The code employs hybrid parallel method of TBB(sheared memory parallel) within each CPU and MPI(distributed memory parallel) for the communication between CPUs. There is no multi-core communication during computation time except diagnose. The diagnose functions in the middle of execution only transfer a limited data and they are usually executed every thousand time steps. Therefore the communication time is negligible . Therefore, the parallel efficiency is close to $100\%$.

\section{Particle simulation results}
\subsection{Test particle simulations}

\begin{figure*}[htbp]
\centering
\begin{minipage}[t]{0.48\textwidth}
\centering
\includegraphics[width=1\linewidth]{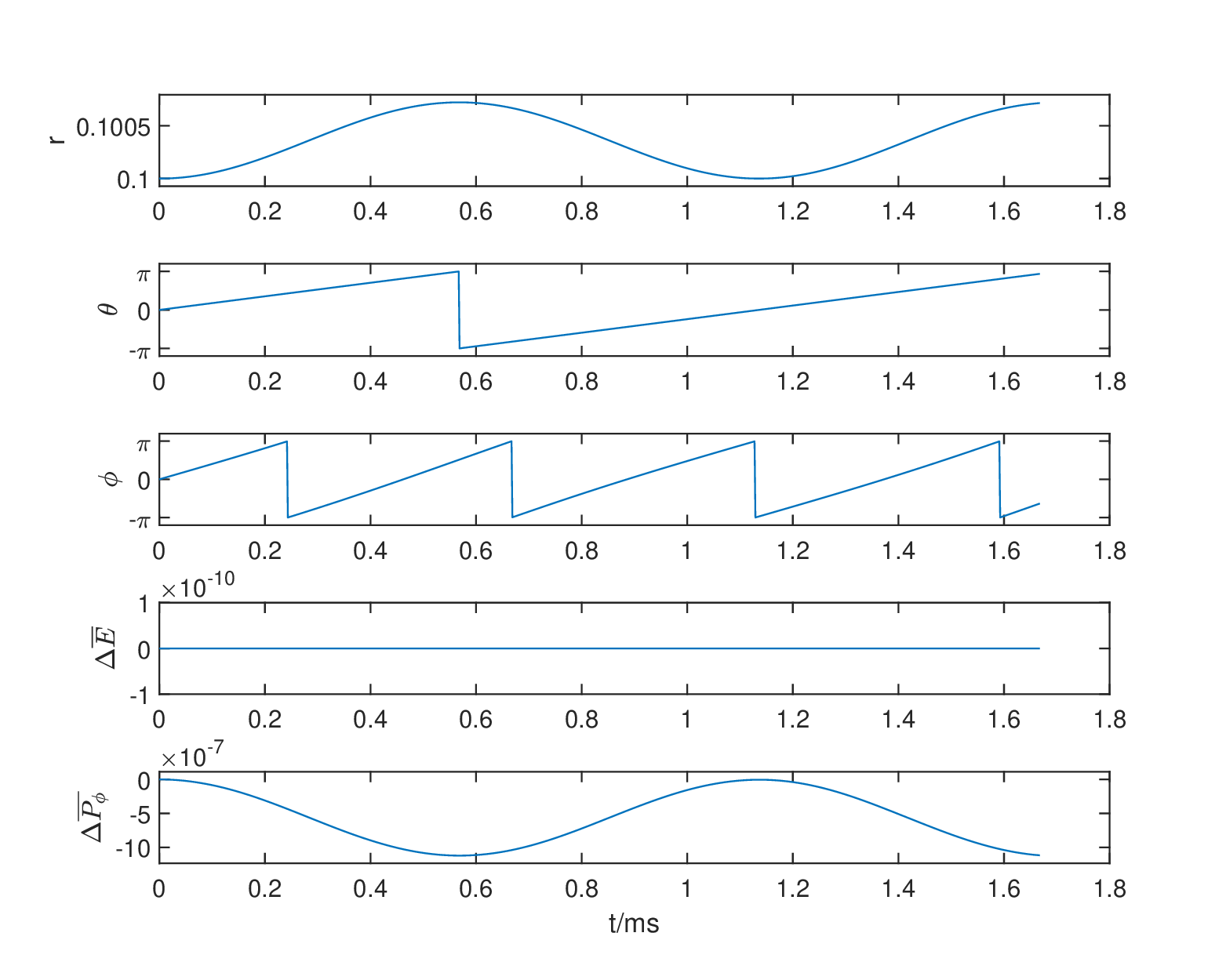}   
\end{minipage}
\begin{minipage}[t]{0.48\textwidth}
\centering
\includegraphics[width=1\linewidth]{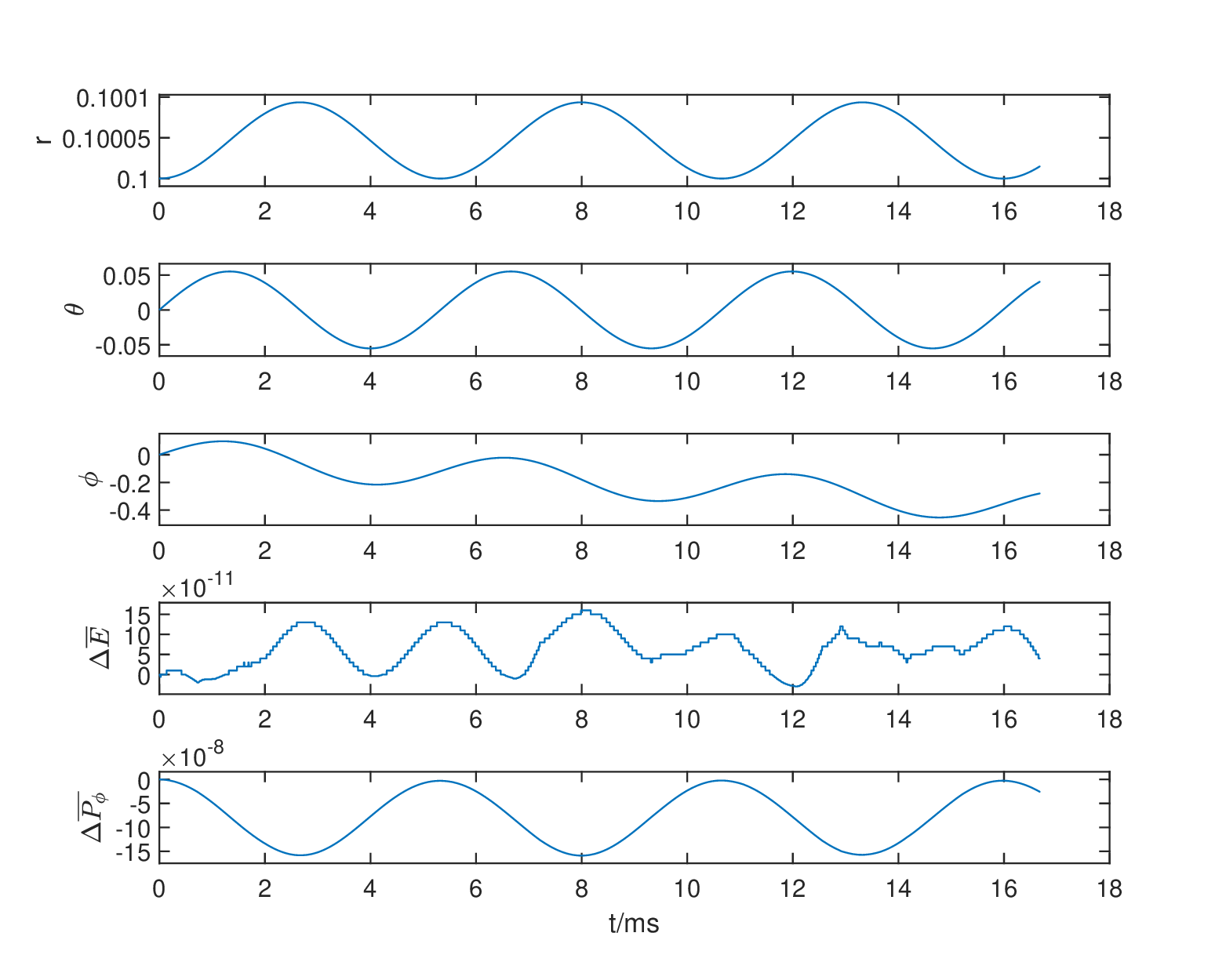}   
\end{minipage}
\caption{Evolution of $r$, $\theta$, $\phi$, the normalized energy change $\Delta \overline{E}=(E-E_0)/E_0$ and the change of the normalized toroidal angular momentum $\Delta \overline{P_\phi}=(P_\phi-P_{\phi0})/P_{\phi0}$ for a passing particle (left panel) and a trapped particle (right panel).}
\label{fig:Tokamakparticle}
\end{figure*}

Here we carry out test particle simulations. First, we consider test particle orbits in a axisymmetric tokamak.
We choose a simple analytic tokamak equilibrium with circular flux surfaces, $B_\phi=-R_0B_0/R$, $B_\theta=rB_0/qR$, and calculate orbits of both passing and trapped particles. The main parameters are the major radius $R=1m$, magnetic field $B_0=1T$, safety factor $q=2.5$ (using a uniform $q$ profile), particle's initial position $r=0.1m$, energy $E=1eV$, particle pitch $v_\parallel/v=1$ for passing particle and $v_\parallel/v=0.01$ for trapped particle. Fig.\ref{fig:Tokamakparticle} shows orbit evolution of radius, poloidal and toroidal angles, normalized energy change and normalized change of the toroidal angular momentum for a passing particle (left panel) and a trapped particle (right panel). The calculated bounce or transit frequency $\omega_b$ and orbit width agree very well with analytic results. The conservations of energy and $P_\phi$ are satisfied accurately. It should be noted that the toroidal angular momentum $P_\phi$ is conserved in tokamaks because of toroidal symmetry.

\begin{figure*}[htbp]
\centering
\subfigure[]{
\begin{minipage}{0.4\linewidth}
\centering
\includegraphics[width=1\linewidth]{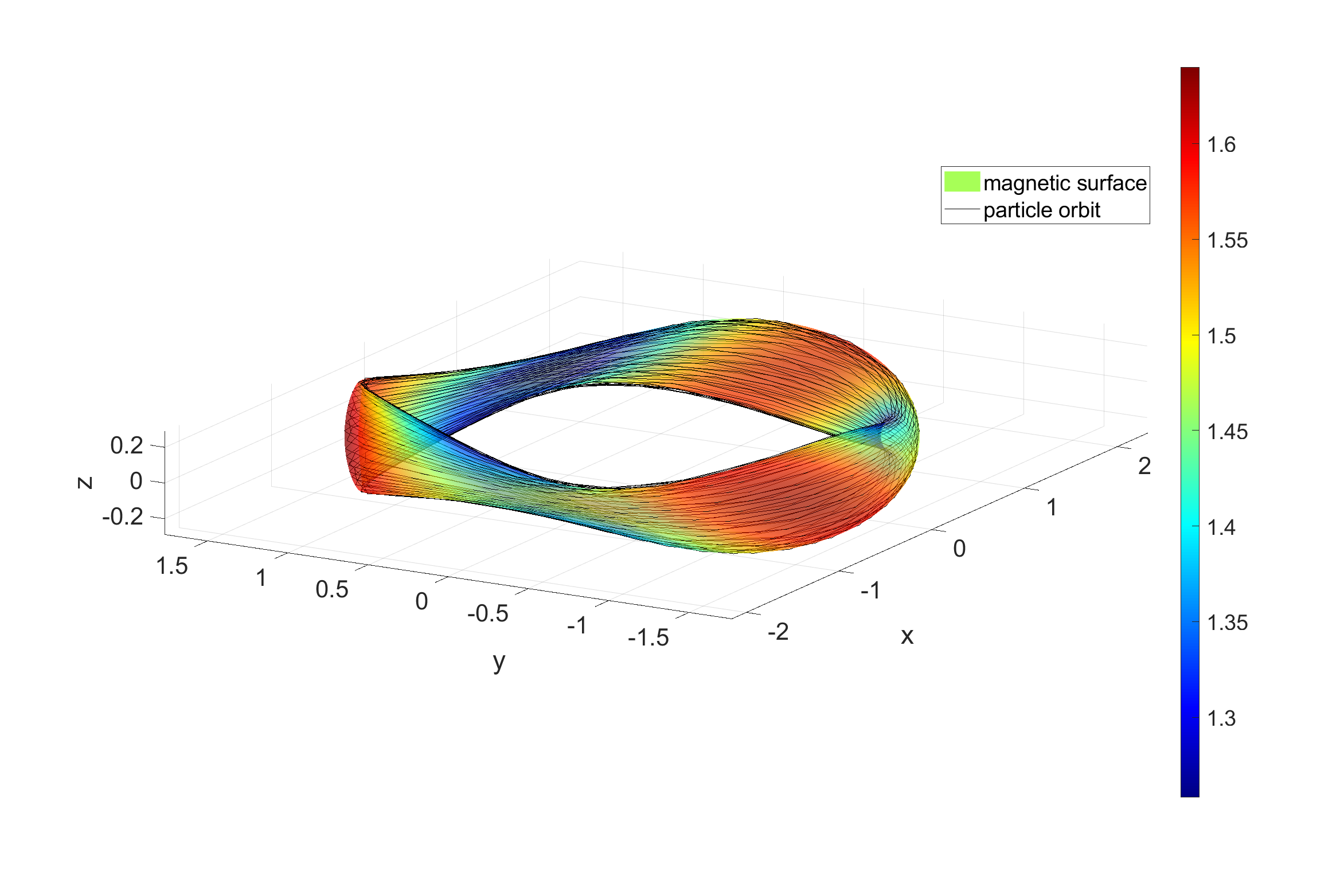}\vspace{4pt}
\includegraphics[width=1\linewidth]{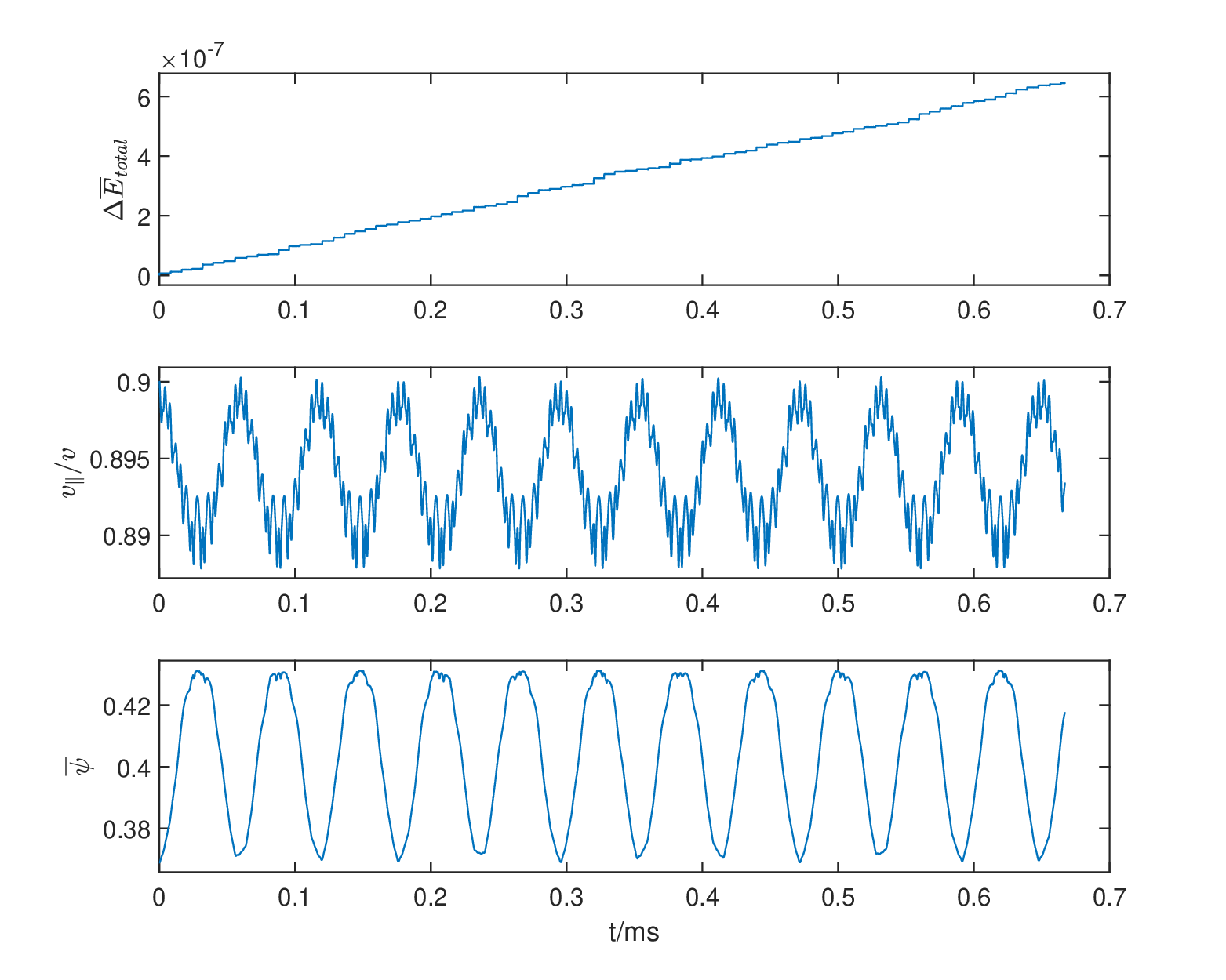}
\end{minipage}%
}
\subfigure[]{
\begin{minipage}{0.4\linewidth}
\centering
\includegraphics[width=1\linewidth]{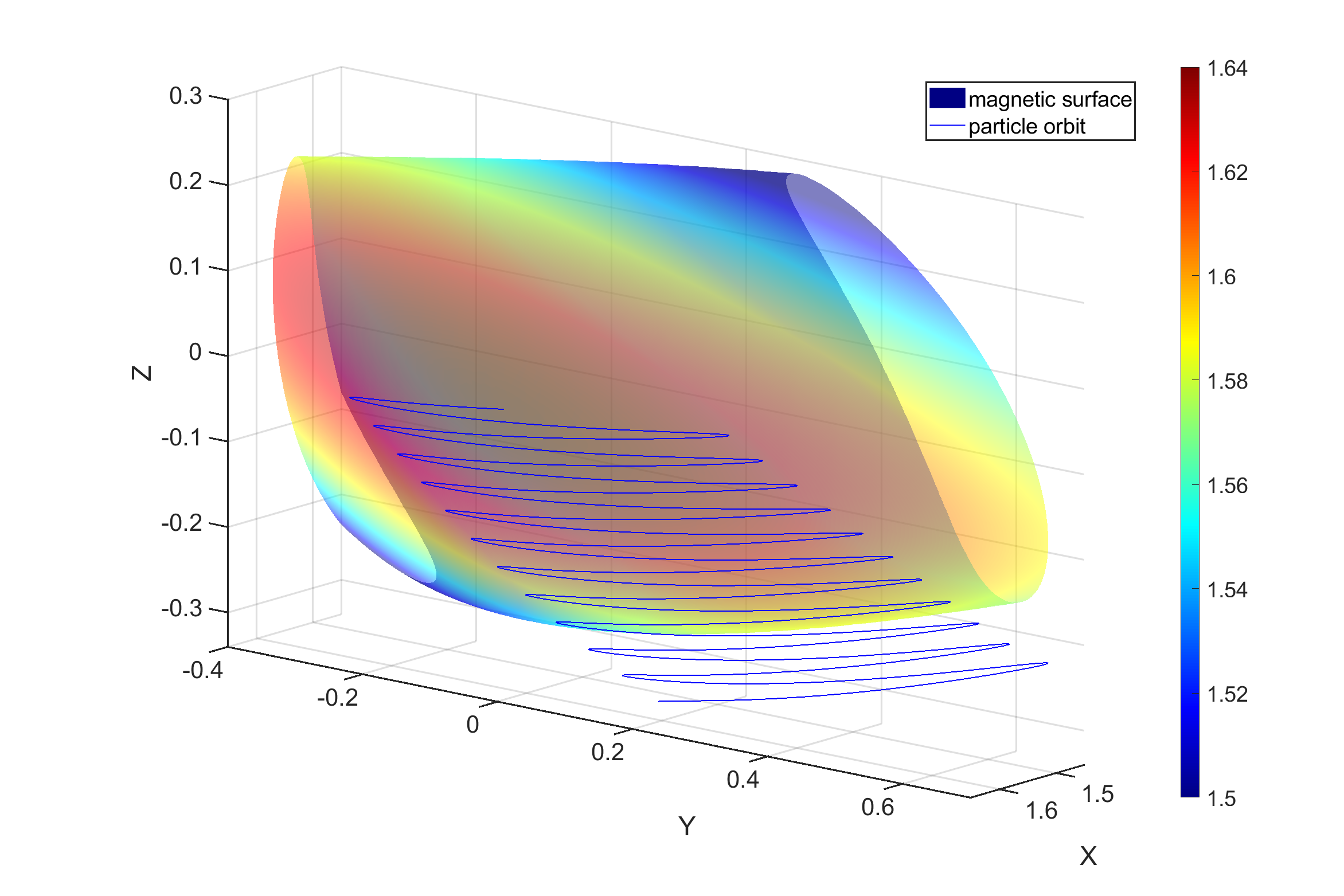}\vspace{4pt}
\includegraphics[width=1\linewidth]{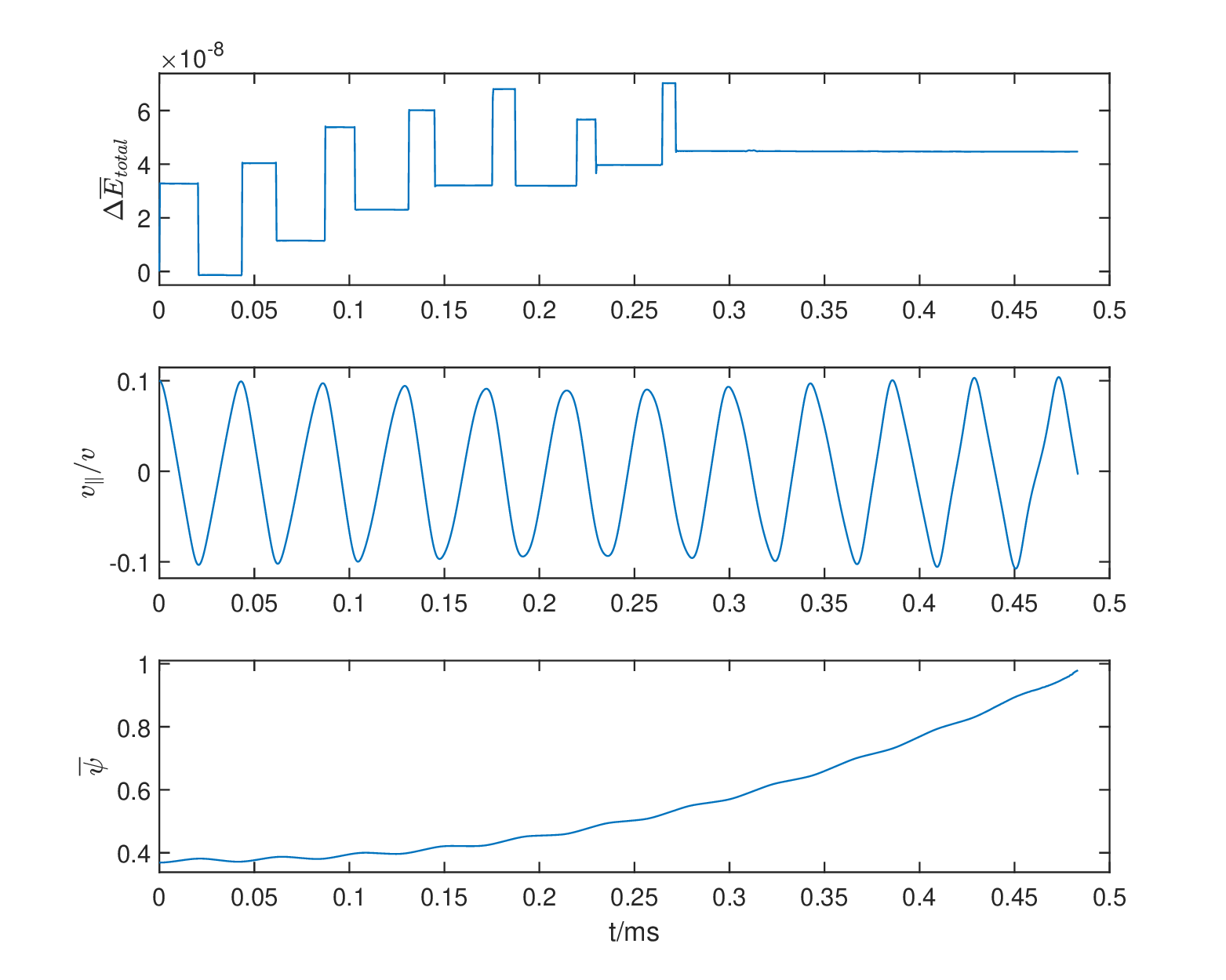}
\end{minipage}
}

\caption{The particle orbit and evolution of the normalized energy change $\Delta \overline{E}=(E-E_0)/E_0$, the normalized parallel velocity $v_\parallel/v$ and the normalized $\psi_n=\psi/\psi_{outermost}$ for a passing particle (left) with $v_\parallel/v=1$  and a trapped particle (right) with $v_\parallel/v=0.1$. Other parameters are $r_0=(1.64,0,0)$, $E=1keV$, time step size $\Delta t=3.34\times10^{-4}ms$ for both of them. }
\label{fig:particleorbit}
\end{figure*}

\begin{figure*}[htbp]
\centering
\subfigure[]{
\begin{minipage}{0.4\linewidth}
\centering
\includegraphics[width=1\linewidth]{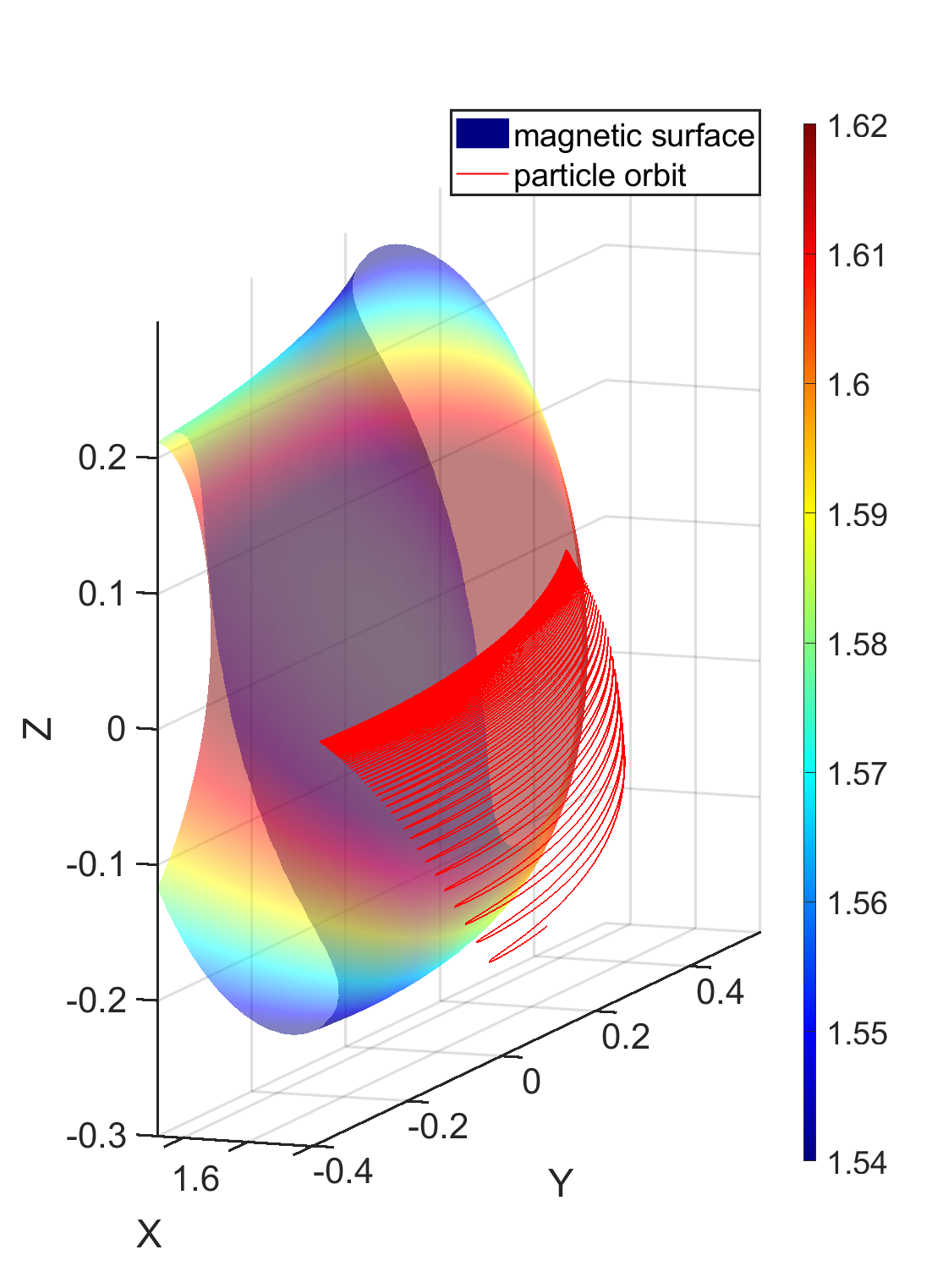}\vspace{4pt}
\includegraphics[width=1\linewidth]{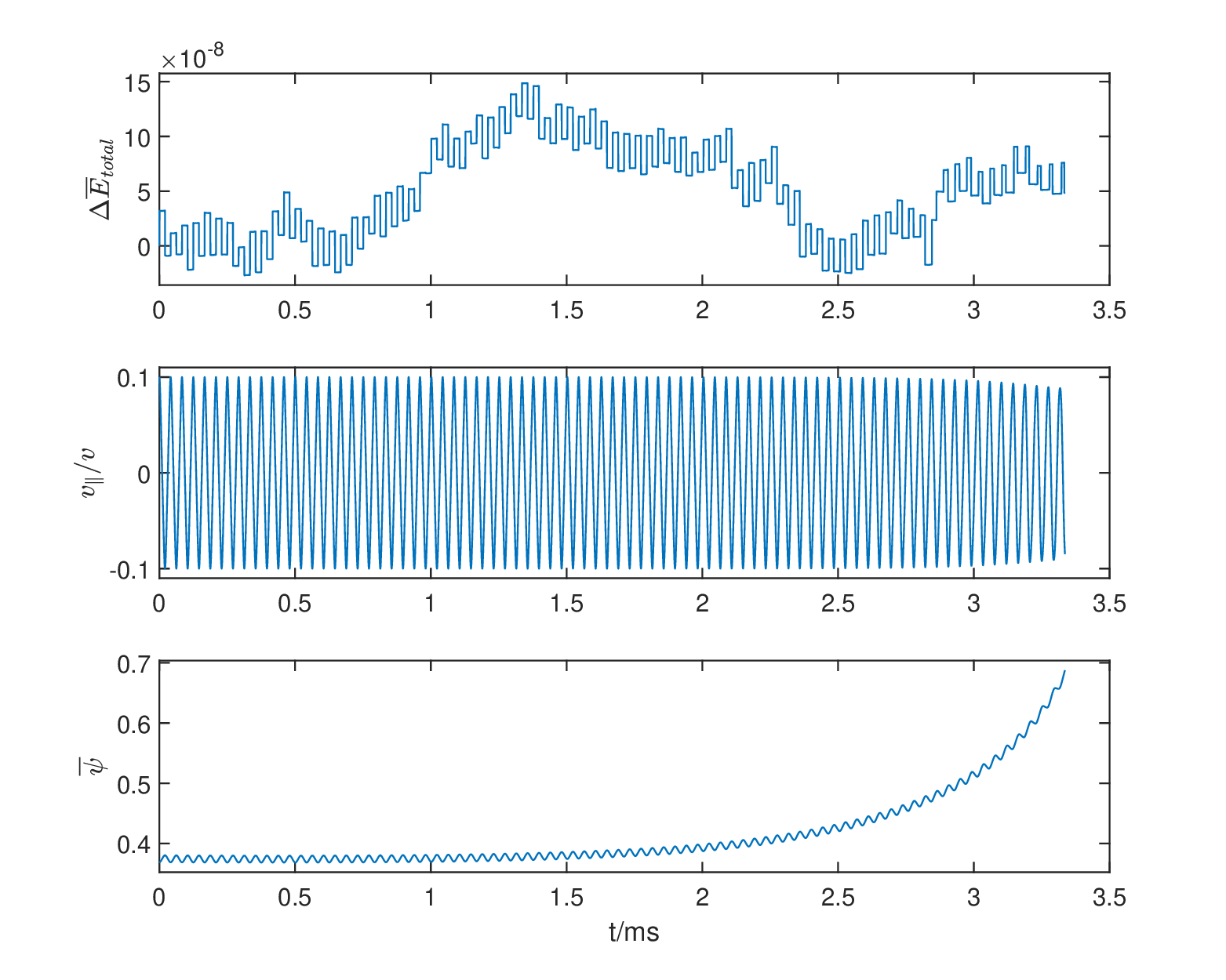}
\end{minipage}%
}
\subfigure[]{
\begin{minipage}{0.4\linewidth}
\centering
\includegraphics[width=1\linewidth]{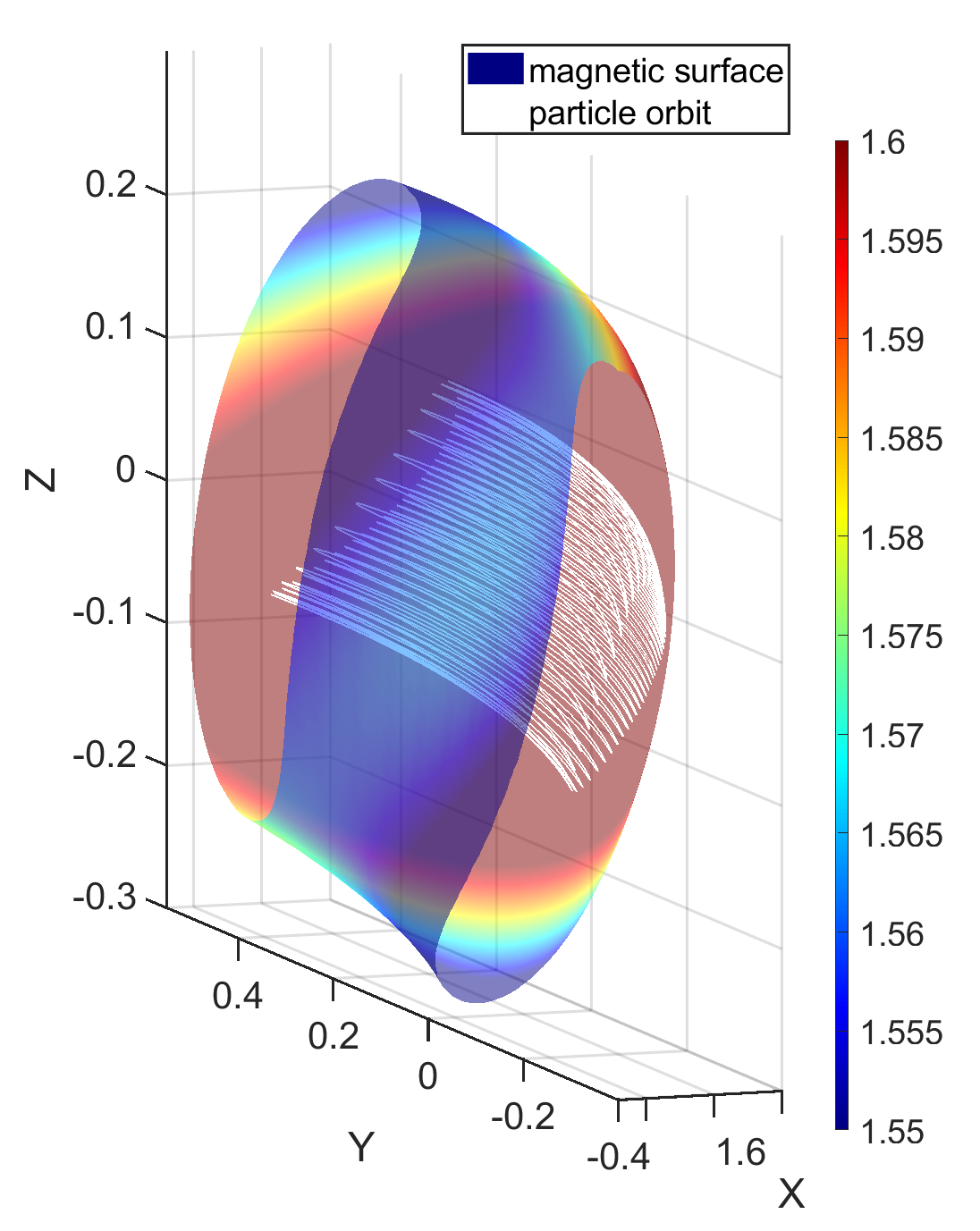}\vspace{4pt}
\includegraphics[width=1\linewidth]{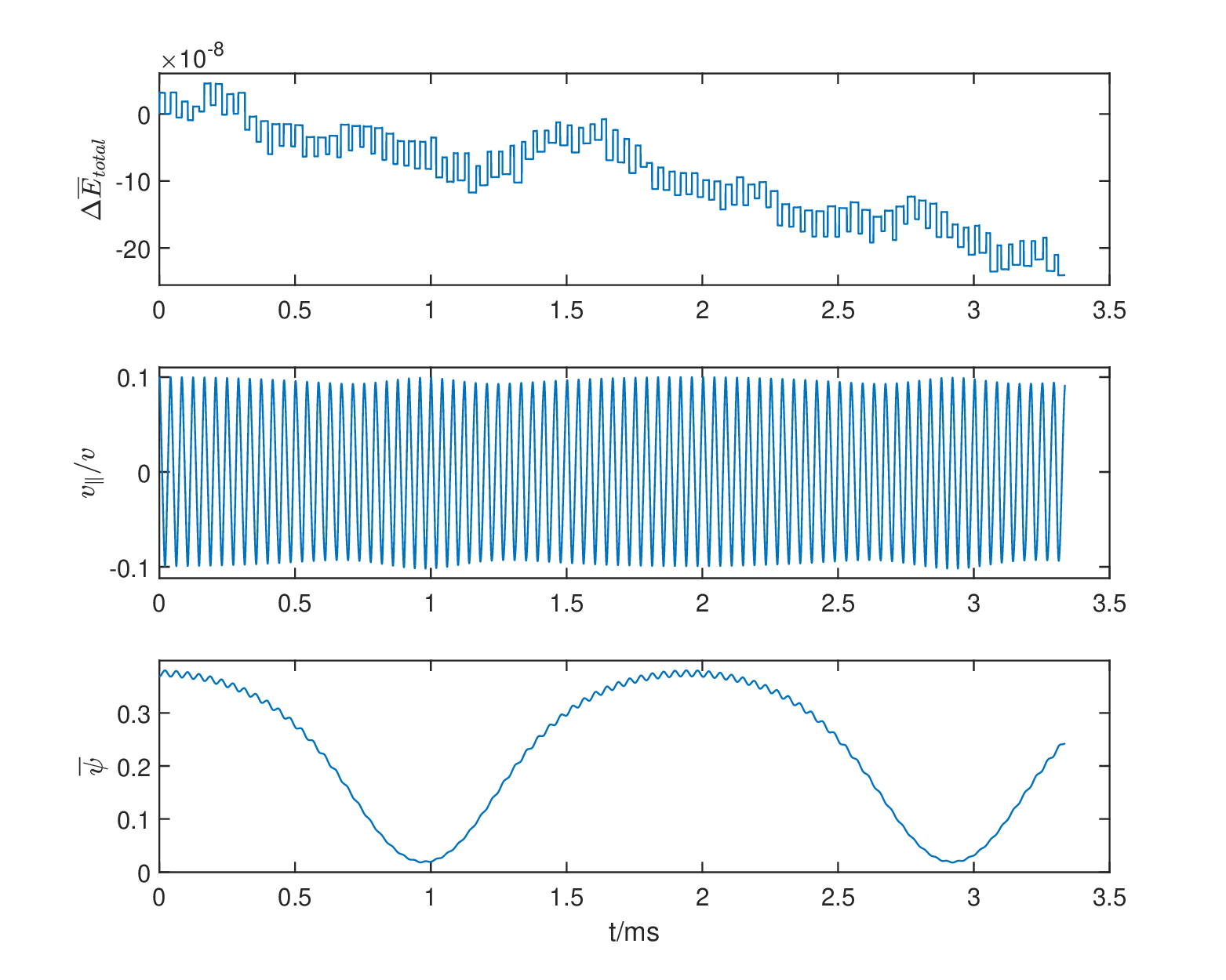}
\end{minipage}
}
\caption{The particle orbit and evolution of the normalized energy change $\Delta \overline{E}=(E-E_0)/E_0$, the normalized parallel velocity $v_\parallel/v$ and the normalized $\psi_n=\psi/\psi_{outermost}$ for a trapped particle with an insufficient electric field (left panel) and a sufficient electric field (right panel) for confining the particle. The total energy is $E_{total} = 1/2mv_\parallel^2+\mu B+q\Phi$.}
\label{fig:particleorbitwithE}
\end{figure*}

We now consider test particle orbits in a stellarator. We choose a typical configuration of NCSX and calculate orbits of both passing and trapped particles. The results are shown in Fig.\ref{fig:particleorbit} without electric field and Fig.\ref{fig:particleorbitwithE} with electric field. Fig.\ref{fig:particleorbit} shows orbit and evolution of energy change, particle pitch and the normalized toroidal flux for a passing particle $v_\parallel/v=1$ (left panel) and a trapped particle $v_\parallel/v=0.1$ (right panel) with the same initial position $r_0=(1.64,0,0)$ and energy $E=1keV$.
The results show that the conservation of total energy is satisfied accurately. It is also shown that the orbit of the passing particle forms a closed surface after many toroidal transits. For the trapped particle without electric field, it drifts downward due to magnetic drift and escapes from the outermost magnetic surface because the averaged drift velocity is not zero due to asymmetry. However, when a radial electric field is present, the effect of the electric field can reduce the averaged drift and the trapped particle can be confined for sufficiently large electric fields as shown in Fig.\ref{fig:particleorbitwithE}. Fig.\ref{fig:particleorbitwithE} shows orbit of a trapped particle for two values of radial electric field in NCSX stellarator with $r_0=(1.64,0,0)$, $E=1keV$, $v_\parallel/v=0.1$. The top left figure corresponds to an electric field not sufficient to confine the particle whereas the top right figure corresponds to an electric field large enough to confine the trapped particle.

\begin{figure}[ht]
\centering
\includegraphics[width=0.5\textwidth]{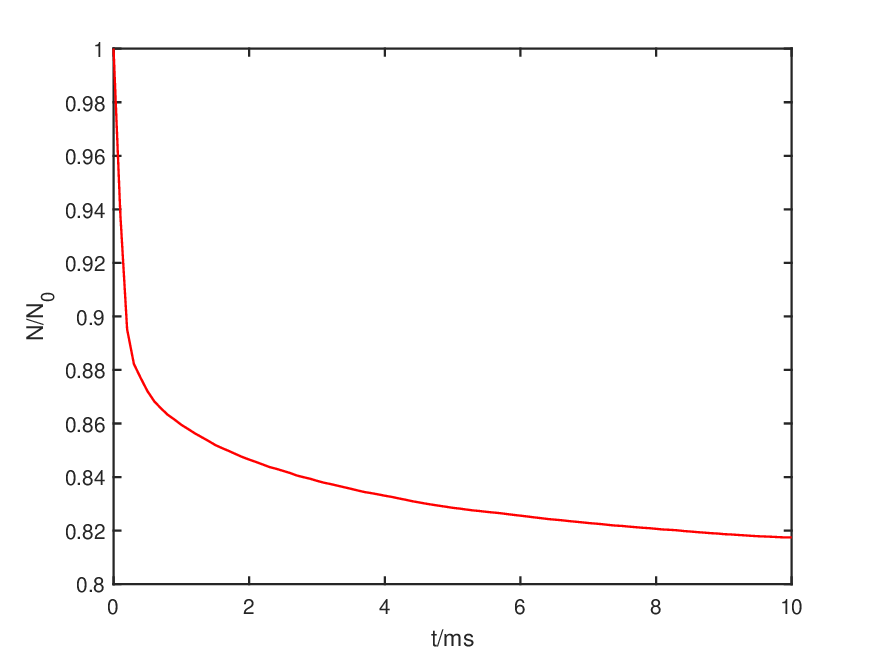}
\caption{Evolution of the normalized number of remaining particles.}
\label{fig:NCSXconstraint}
\end{figure}

To investigate the alpha particles losses in future stellarator reactors, we choose NCSX stellarator with $B_0 = 2T$, $r_{minor}=0.33m$, but the normalized Larmor radius $\rho/r_{minor}$ is set to a typical fusion reactor value of $0.02$, so the energy of alpha particles is chosen to be $10KeV$. We load $10^5$ particles uniformly at one magnetic surface ($\psi_n = 0.4$) with same energy and random pitch angle at the beginning of simulation and simulate $10ms$ without electric field and collision. The boundary of simulation is chosen to be $\psi_n = 0.98$. The normalized number of confined particles is shown in Fig \ref{fig:NCSXconstraint}. We observe that $12\%$ of particles are lost in a short time ($0.3ms$), and $18\%$ of particles are lost in $10ms$.

\subsection{Simulation of neoclassical transport}
The radial particle diffusive flux is given by
\begin{equation}\label{3.6}
  \Gamma_\psi=-D_{\psi\psi}\frac{\partial n}{\partial\psi}
\end{equation}
where the diffusion coefficient $D_{\psi\psi}$ is a function of $\psi$. The transport coefficient is calculated by
\begin{equation}\label{3.7}
  D_{\psi\psi} = \frac{1}{2} \frac{\partial \overline{(\Delta\psi)^2}}{\partial t}
\end{equation}
where $\overline{(\Delta\psi)^2}$ is the average of the square of the radial deviation $\Delta\psi$. It should be noted that $\overline{(\Delta\psi)^2}$ increases linearly with time for diffusive transport. We can also calculate $D_{rr}$ by replacing $\psi$ with $r=\sqrt{\overline{\psi}}r_0$ where $r_0$ is a normalizing radial length.

\begin{figure}[htbp]
\centering
\subfigure[]{
\begin{minipage}[t]{0.48\textwidth}
\centering
\includegraphics[width=1\linewidth]{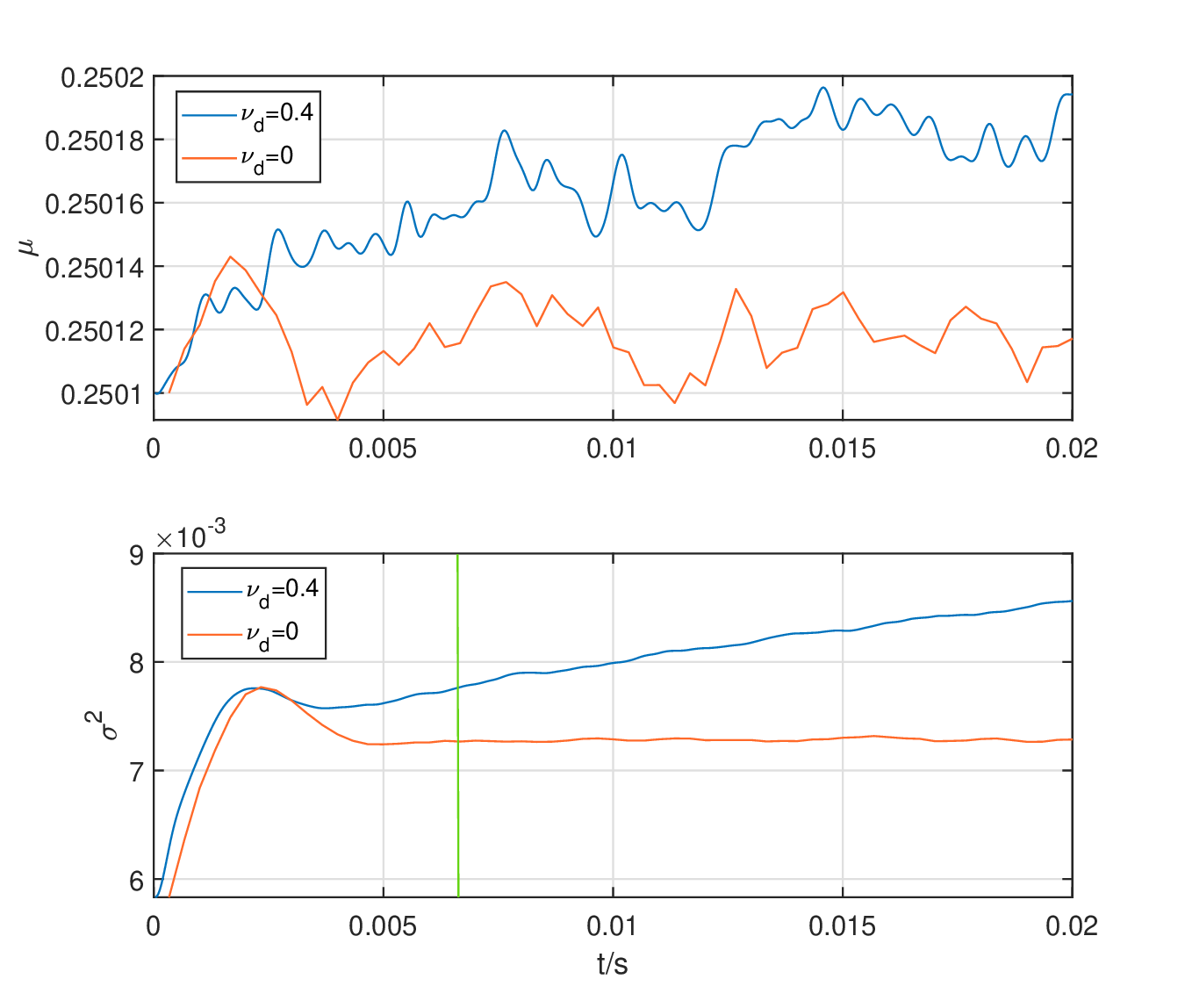}
\end{minipage}%
}
\subfigure[]{
\begin{minipage}[t]{0.48\textwidth}
\centering
\includegraphics[width=1\linewidth]{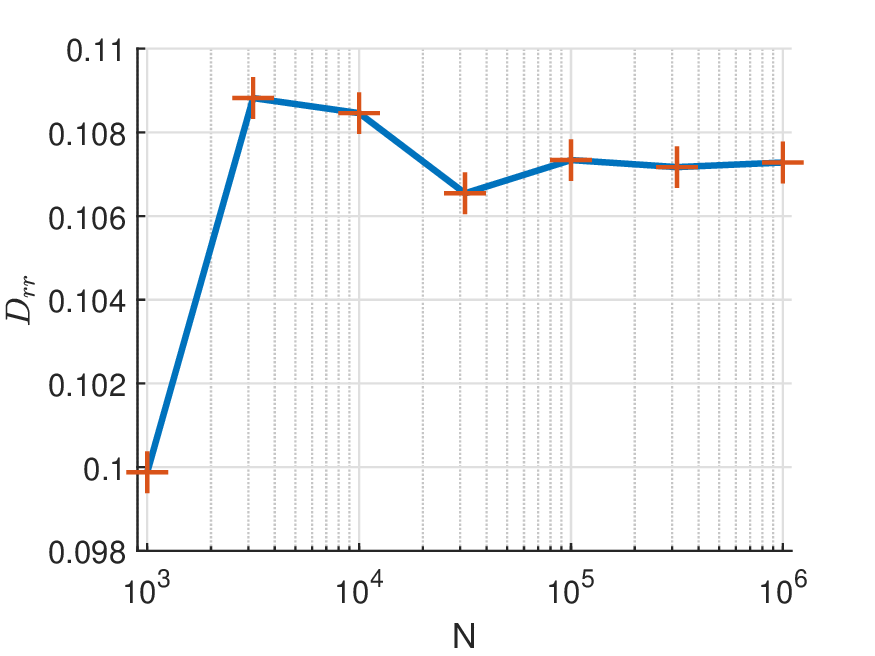}
\end{minipage}
}
\caption{(a) shows the gaussian fitting parameter $\mu$(expectation), $\sigma^2$(variance) of $\psi$. (b) shows the convergence of $Drr$ with respect to particle number N.}
\label{fig:Drr_fitting}
\end{figure}

Particles are loaded uniformly at one magnetic surface at the beginning of simulation ($\overline{\Delta\psi^2}_0=0$). The evolution of the Gaussian fitting parameter $\mu$ (expectation) and $\sigma$ (standard deviation) are shown in Fig.\ref{fig:Drr_fitting}(a). We observe that the center of distribution changes little throughout the simulation. The effect of orbit width causes increase of $\sigma$ in short time initially, and the distribution of $\psi$ is not exactly an Gaussion due to the different orbit width for inner side and outer side. After the time indicated by the green line, the collision dominates the increase of $\sigma$. The distribution tends to be a Gaussion and the slope converges as time increases. The accuracy of $D_{rr}$ and $D_{\psi\psi}$ rely heavily on the number of simulation particles N as shown in Fig.\ref{fig:Drr_fitting}(b). When $N>10^5$, the transport coefficient is well converged.

In order to benchmark transport coefficient $D_{rr}$, we use the tokamak analytical result\cite{lin1995gyrokinetic} of neoclassical transport in both small collision limit $D_l$ and collisional $D_c$ limit given below
\begin{flalign}
  D_l &= \frac{3}{8}I_1\nu \rho^2 \frac{q^2}{\epsilon^2} \label{4.1} \\
  D_c &= \nu q^2 \rho^2 \label{4.2}
\end{flalign}
where $\rho$ is the thermal gyroradius ($\rho=mv_{th}/eB_0$), q is safety factor and $\epsilon$ is inverse aspect ratio, and to the lowest order in $\epsilon$,
\begin{equation}\label{4.3}
  I_1=1.38\sqrt{2\epsilon}
\end{equation}

\begin{figure}[ht]
\centering
\includegraphics[width=1\linewidth]{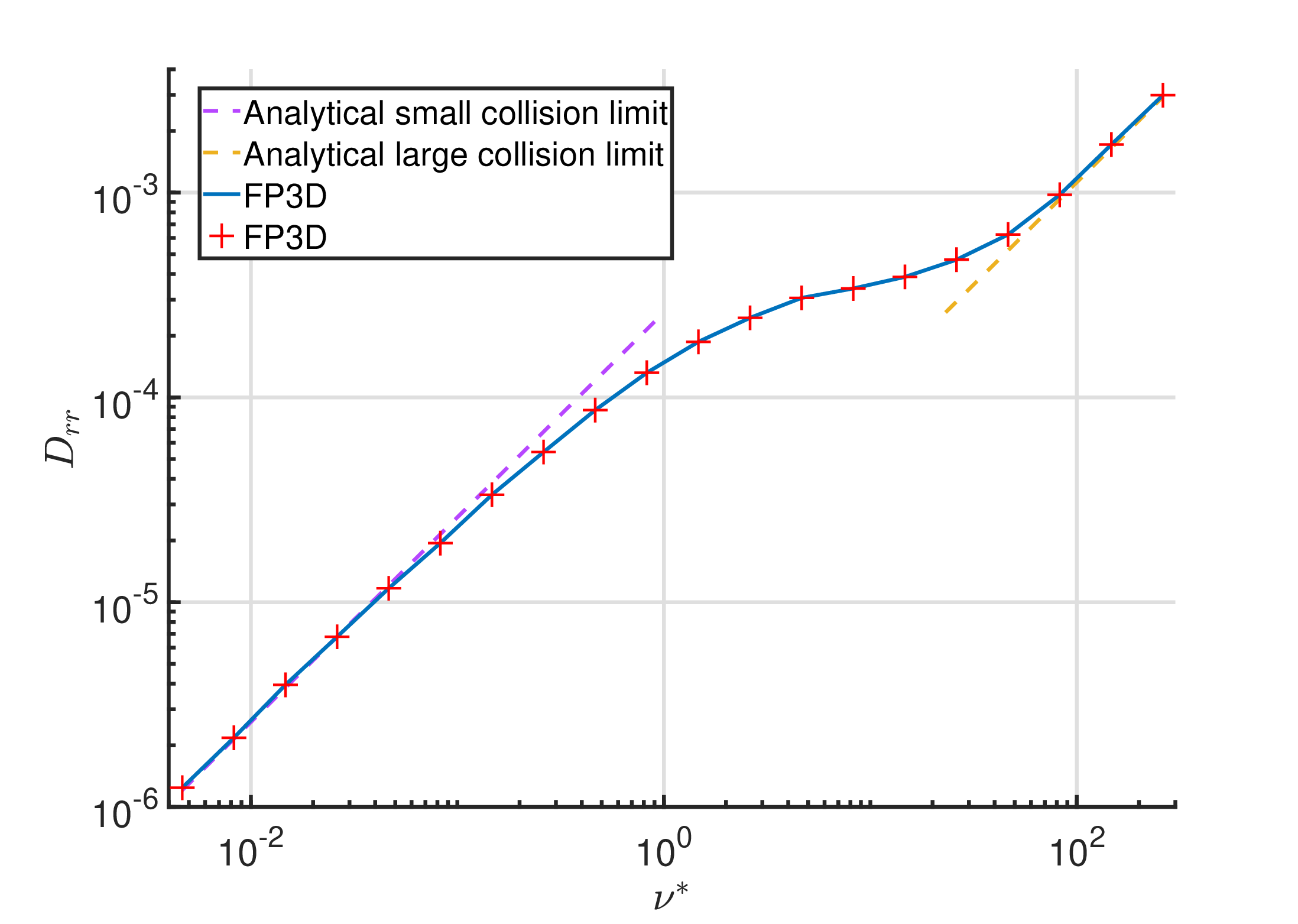}
\caption{Comparison of diffusion coefficients as a function of collision frequency. Red crosses are obtained from FP3D while purple line and orange line correspond to small collision frequency limit and large collision frequency  limit of analytic theory. Key parameters used are $R_0=1m$, $B_0=1T$,$\epsilon=0.1$ and $q=2.5$, and the number of particle is $10^5$.}
\label{fig:Drr_Tokamak}
\end{figure}

Fig.\ref{fig:Drr_Tokamak} shows the dependence of particle diffusion coefficient on the effective collision frequency $\nu^*$ with $\nu^*=\epsilon^{-3/2}\nu\sqrt{2}qR_0/v_{th}$. We observe that the simulation results coincide with the analytic results. The simulation takes less than 1 min for simulating 10000 steps with $10^5$ particles and 160 CPU for each case.


\begin{figure}[ht]
\centering
\includegraphics[width=1\linewidth]{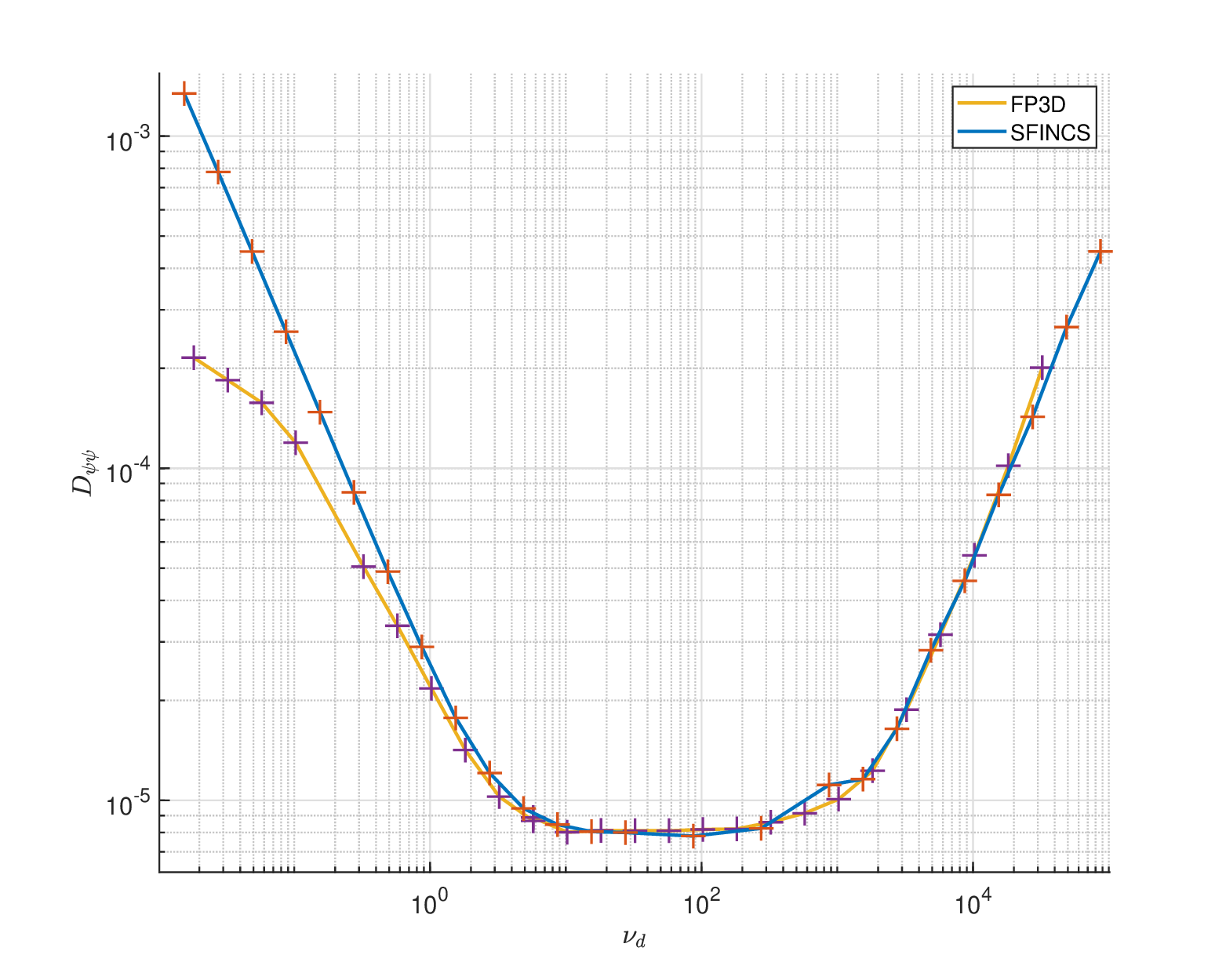}
\caption{Comparison of diffusion coefficient $D_{\psi\psi}$ obtained from FP3D (orange line) and SFINCS (blue line) as a function of collision frequency in NCSX. The results are obtained with a single energy of $E=1eV$ and an isotropic distribution in pitch angle.}
\label{fig:Drr_NCSX}
\end{figure}

For stellarators, we compare our results with those obtained from the Stellarator Fokker-Planck Iterative Neoclassical Conservative Solver (SFINCS)\cite{landreman2014comparison}. The vacuum equilibrium was obtained from VMEC. For simplicity, electric field is assumed to be zero. The temperature profile is chosen to be uniform. Thus, the neoclassical transport for a single ion species is driven by density gradient in this case.

The results from FP3D and SFINCS are shown in Fig.\ref{fig:Drr_NCSX}. The simulation results are consistent with SFINCS's except in low $\nu_d$ regime. The discrepancy is probably due to the deviation from Gaussion distribution in $\psi$ in the low collision regime, so the $\overline{\Delta\psi^2}$ is inaccurate. On the other hand, the distribution becomes a Gaussian one in short time due to the collision effect in the middle and high $\nu_d$ regime. So the results are consistent with those of SFINCS.

The simulations are more expensive in above stellarator case. It takes about 1 hour with 160 cores on Intel Xeon Gold 6148 CPU for simulating 15000 steps with $10^5$ particles.

\subsection{Simulation of ripple losses}
In a real tokamak, finite number of toroidal field coils breaks the toroidal symmetry in magnetic field and induces field ripple in toroidal direction\cite{PhysRevLett.47.647}. This asymmetry results in losses of fast ions, including the ripple well losses and the ripple stochastic losses. The magnetic field ripple is given by $\delta\mathbf{B}=\delta B^{rip}_{\phi}\nabla\phi$. The ripple perturbation $B^{rip}_{\phi}$ can be written as\cite{yingfeng2021simulations}
\begin{equation}
  \delta B^{rip}_{\phi}=-B_0 R_0 \delta(R,Z)cos(N\phi)
\end{equation}
where N is the number of the toroidal field coils and $\delta$ is the normalized ripple amplitude.

\begin{figure}[htbp]
	\centering
	\begin{minipage}[b]{0.45\linewidth}
		\centering
        \includegraphics[width=1\linewidth]{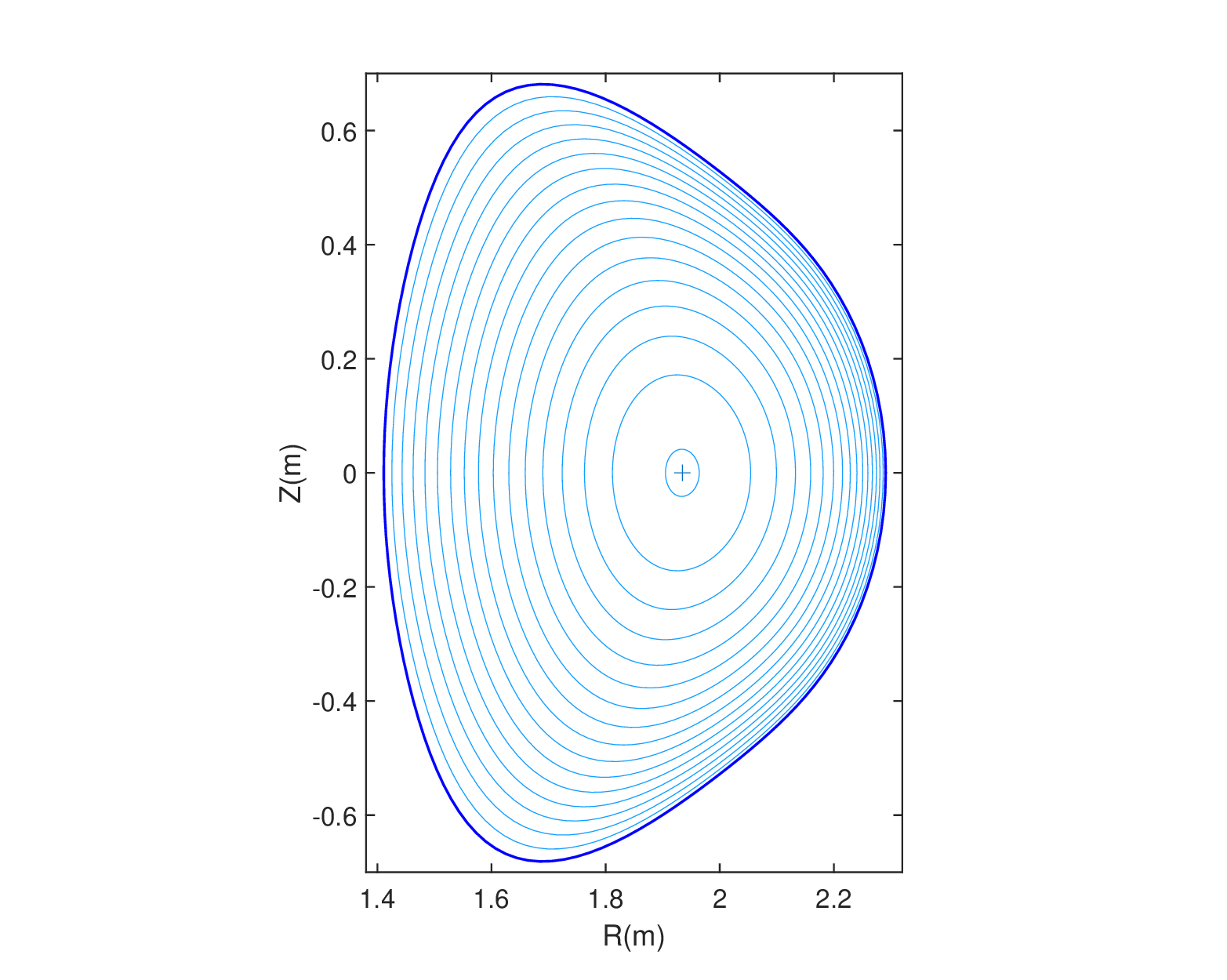}
        \caption{The Poincare section of magnetic field line (blue lines) and the shape of the VMEC boundary (magenta line) on EAST.}
        \label{fig:EAST_mgf}
	\end{minipage}
	\qquad
	\begin{minipage}[b]{0.45\linewidth}
		\centering
        \includegraphics[width=1\linewidth]{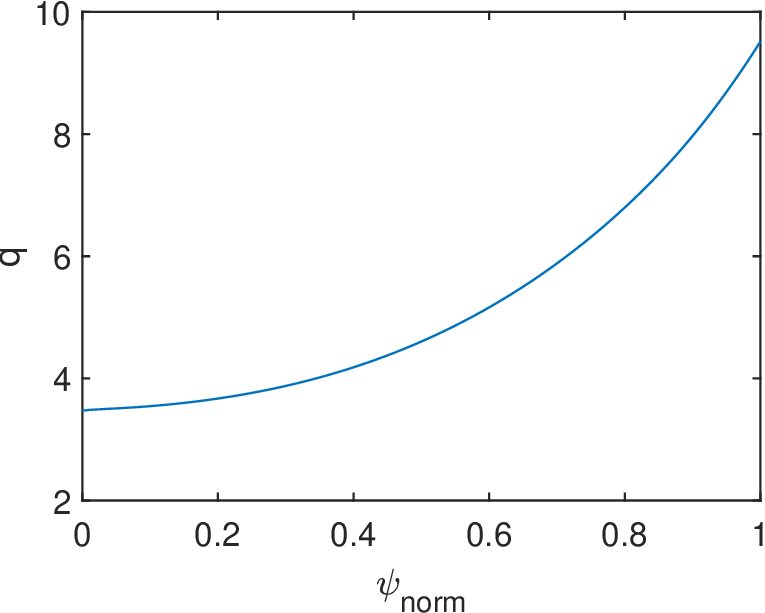}
        \caption{The safety factor profile of EAST.}
        \label{fig:EAST_q}
	\end{minipage}
\end{figure}

We use parameters and profiles for a discharge of the Experimental Advanced Superconducting Tokamak (EAST)\cite{yingfeng2021simulations}. The magnetic equilibrium is calculated by VMEC. The main parameters are shown in Fig.\ref{fig:EAST_mgf} and \ref{fig:EAST_q}. The ripple amplitude $\delta$ can be fitted by an analytic function, expressed as\cite{yingfeng2021simulations}
\begin{equation}
  \delta(R,Z)=\delta_0 exp\left\{[(R-R_{rip})^2+b_{rip}Z^2]^{1/2}/w_{rip}\right\}
\end{equation}
For the EAST tokamak, the parameters related to the toroidal field ripple are given as $N= 16, \delta_0=1.267\times 10^{-4}, R_{rip}=1.714-0.181Z^2,b_{rip}=0.267, w_{rip}=0.149m$.\cite{yingfeng2021simulations}

\begin{figure}
\centering
\subfigure[]{
	\begin{minipage}[t]{0.45\linewidth}
		\centering
		 \includegraphics[width=\linewidth]{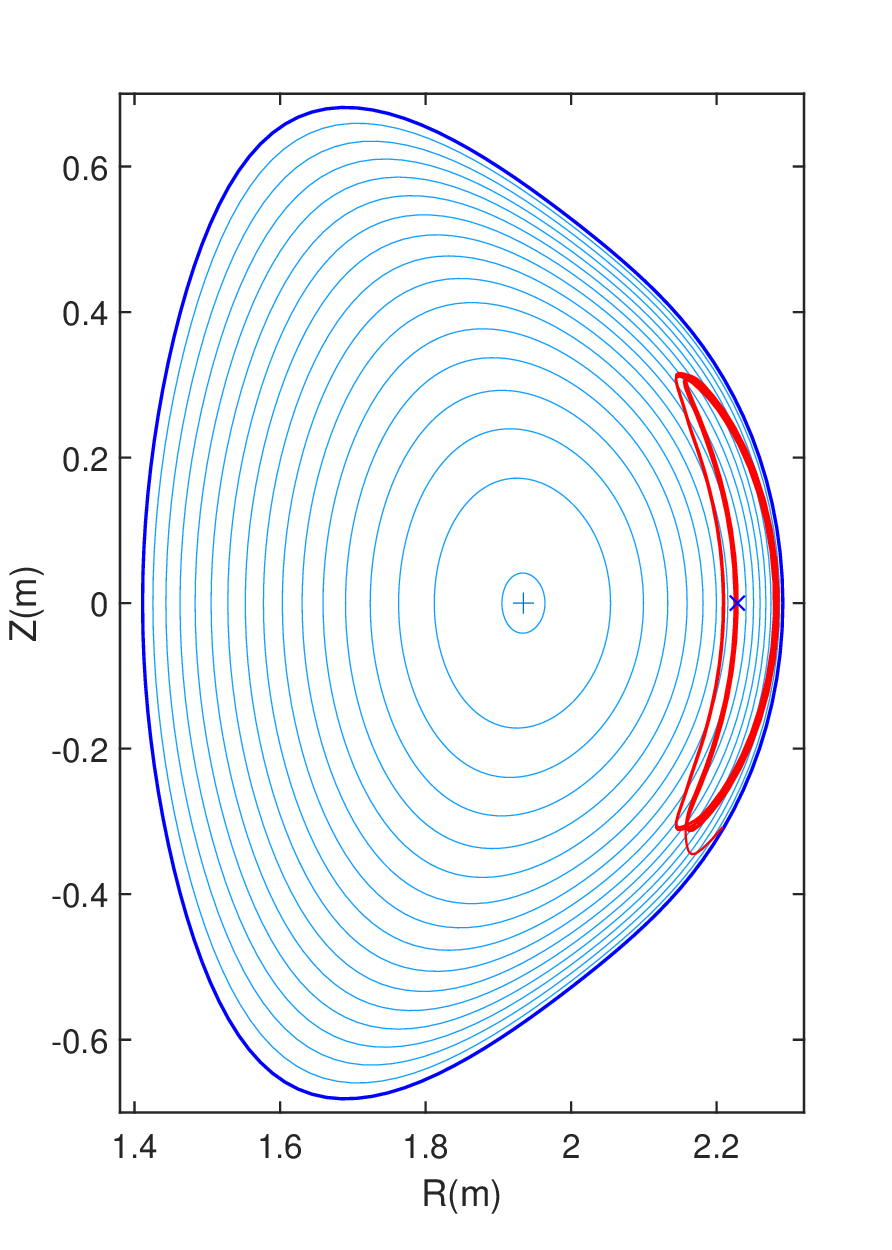}
	\end{minipage}
}
\subfigure[]{
	\begin{minipage}[t]{0.45\linewidth}
		\centering
		\includegraphics[width=\linewidth]{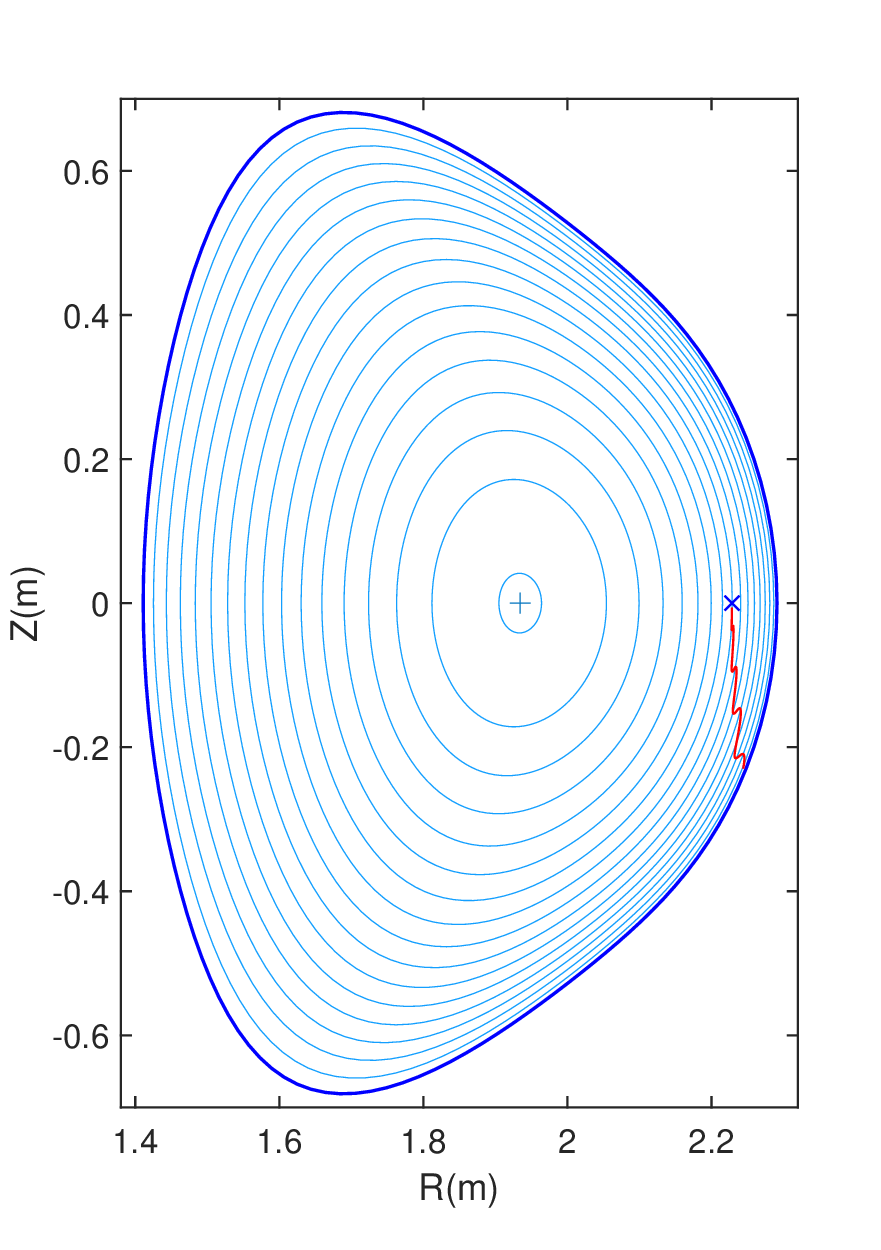}
	\end{minipage}
}
	\caption{Orbits (red line) of two trapped protons with energy 50keV in the EAST. The initial pitch is (a)$v_\parallel / v=$0.2 and (b) 0.05. The initial radial position is $\hat{\psi}=0.5, \theta=0, \phi=0$. The dark blue cross symbols denote the initial positions of a proton.}
	\label{fig:EAST_trapped}
\end{figure}

\begin{figure}
\centering
\subfigure[]{
	\begin{minipage}[t]{0.45\linewidth}
		\centering
		 \includegraphics[width=\linewidth]{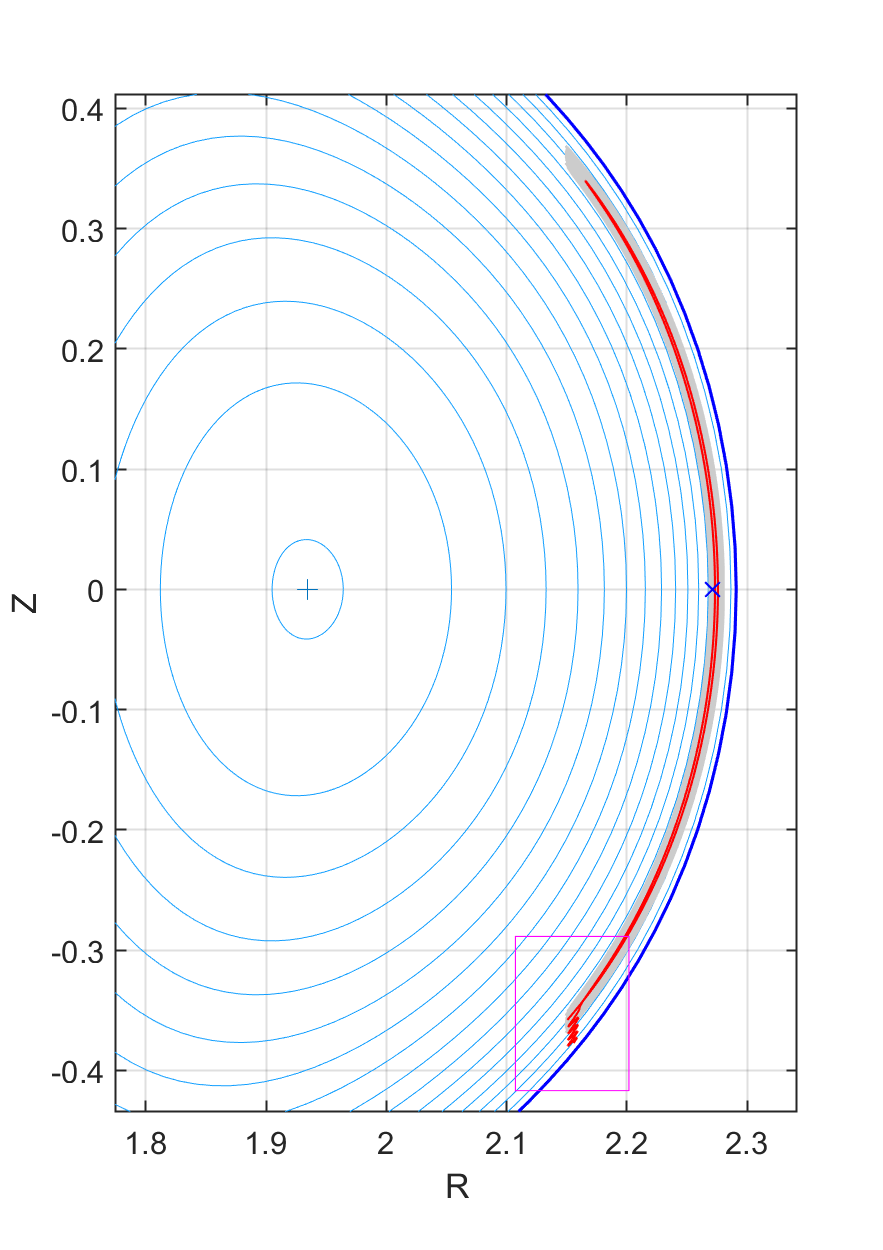}
	\end{minipage}
}
\subfigure[]{
	\begin{minipage}[t]{0.45\linewidth}
		\centering
		\includegraphics[width=\linewidth]{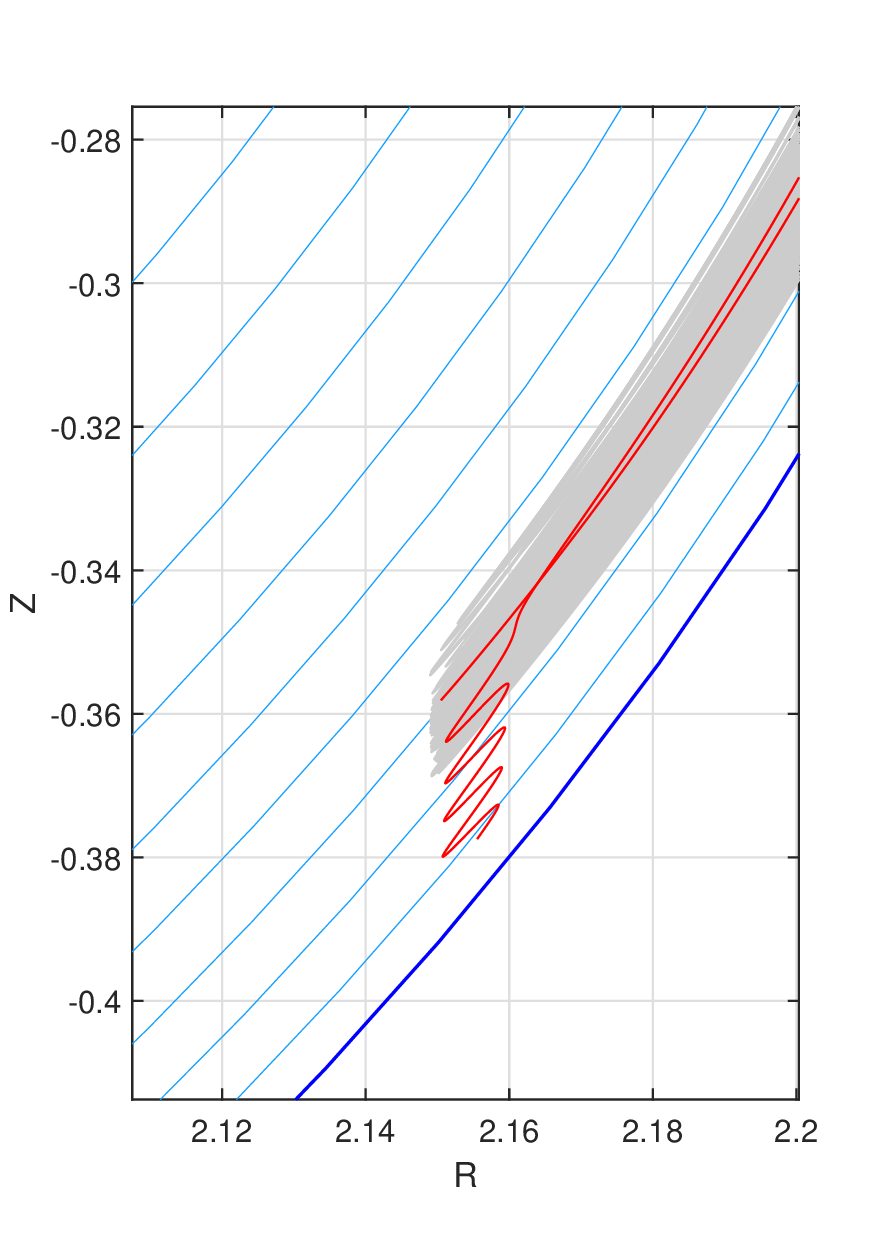}
	\end{minipage}
}
\caption{(a) The whole orbit (gray line) and the final orbit (red line) of a trapped protons with energy 1keV in the EAST tokamak. The initial radial position is $\hat{\psi}=0.8, \theta=0, \phi=0$. (b) An expanded figure of the pink area in (a). And the particle escape below at $Z=-0.38$.}
\label{fig:EAST_convert}
\end{figure}

Fig.\ref{fig:EAST_trapped}(a) shows that the particle orbit with the initial pitch $v_\parallel /v=0.2$ finally hits the boundary after a long time because of ripple stochastic loss. On the other hand, from Fig.\ref{fig:EAST_trapped}(b), we can see that the particle with a smaller pitch $v_\parallel /v=0.05$ crosses the boundary vertically below the initial position in a very short time. This fast loss is due to the ripple well trapping. These results are consistent with the results of Yingfeng Xu\cite{yingfeng2021simulations}. In Fig.\ref{fig:EAST_convert}, it is very interesting that some particles convert from barely trapped to ripple trapped after many bounces (arrow pointing position) and then escape from below immediately.

A key parameter scanning shows region of ripple losses. The particles' initial radial points are $\overline{\psi}=0.3$, $0.4$ and $0.5$ at the mid-plane, respectively and their toroidal angles $\phi=0$ are at the weakest magnetic field section.

\begin{figure}[hbt!]
        \centering
        \subfigure[]{
            \begin{minipage}[b]{0.5\textwidth}
            \includegraphics[width=\textwidth]{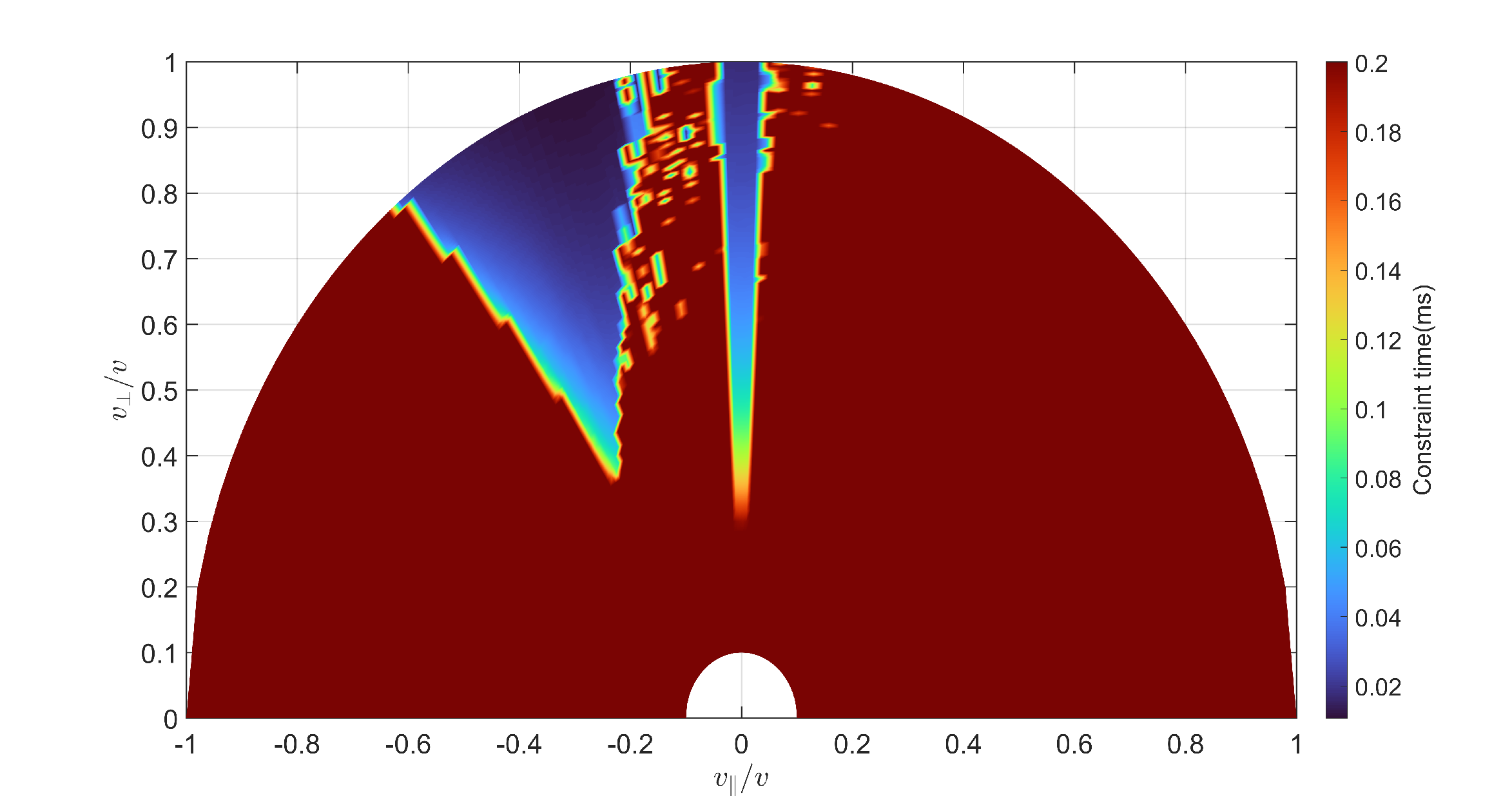}
            \end{minipage}
            }

        \subfigure[]{
            \begin{minipage}[b]{0.5\textwidth}
            \includegraphics[width=\textwidth]{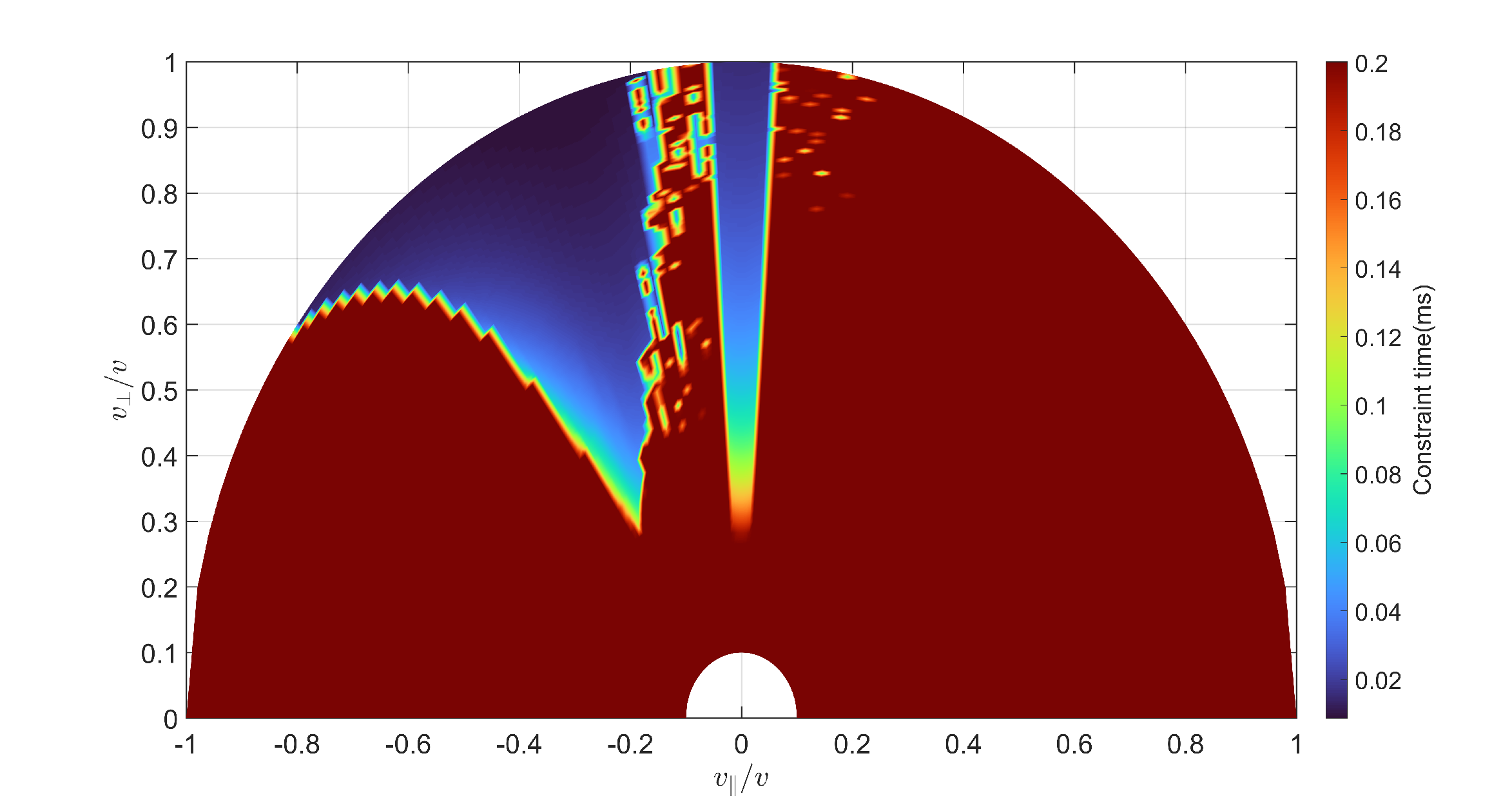}
            \end{minipage}
            }

        \subfigure[]{
            \begin{minipage}[b]{0.5\textwidth}
            \includegraphics[width=\textwidth]{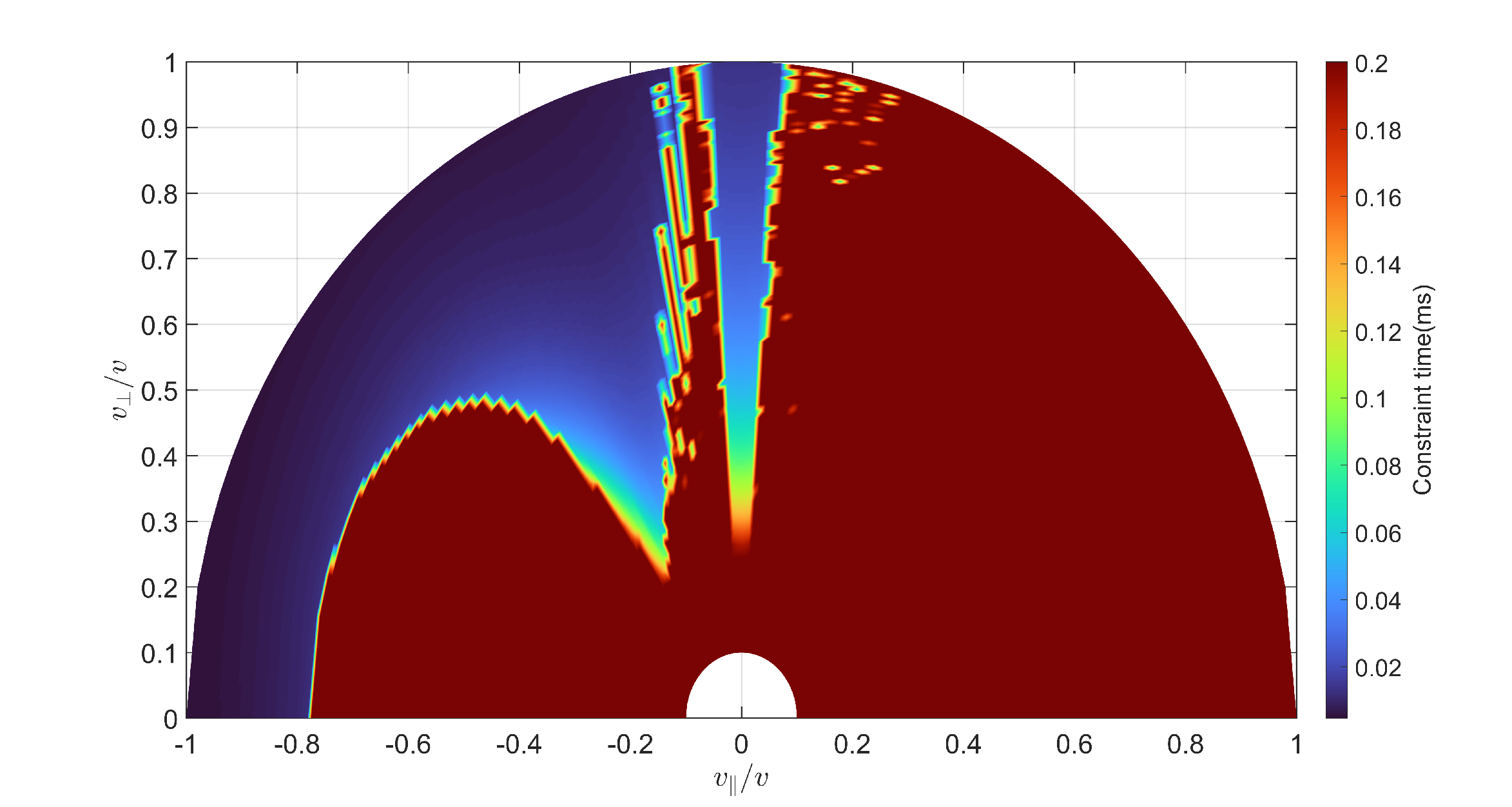}
            \end{minipage}
            }
        \caption{Confinement time in velocity space for collisionless protons with the starting radial position of (a) $\psi$ = 0.3, (b) 0.4 and (c) 0.5. Both initial $\theta$ and $\phi$ are set at 0.}
        \label{fig:EAST_Constraint_time}
\end{figure}

Fig.\ref{fig:EAST_Constraint_time} shows the confinement time of collisionless protons in different energies and initial pitch angles. The confinement time refers to the time during which particles are confined inside the boundary($\overline{\psi}=0.98$) during the maximum simulation time of 0.2ms.
There is a narrow loss cone in $v_\parallel=0$ region because of ripple-trapping. The other loss regions are caused by stochastic ripple loss. This result is consistent with the previous result of K. TANI \cite{tani1983ripple}.



\section{Summary and Conclusion}

An efficient test particle code FP3D is developed. The main functions of the code include (1) magnetic field calculation from 3D coils, (2) field line tracing, (3) test particle simulations in 3D magnetic fields. In the first function, the magnetic field can also be obtained from equilibrium codes or analytic models. In the second function, FP3D traces field lines to determine magnetic surfaces, magnetic axis, and calculate rotational transform as well as flux coordinates $(\psi,\theta,\phi)$. In the last function, it calculates test particle orbits and neoclassical transport coefficient in 3D magnetic fields. The code can also be used to calculate the distribution of lost particles. The code has been verified comprehensively through consistency check, conservation of particle energy as well as benchmark with analytic theory and other codes.

The code has been applied successfully to calculate the neoclassical transport coefficient with results in agreement with those of the SFINCS code. The code has also been applied to calculate ripple losses in the EAST experiment with results consistent with previous work.

In conclusion, FP3D is a comprehensive code for field calculation and test particle simulations in 3D magnetic fields. It is an efficient parallel code with advanced solver. The abstract code interface make it easy to add new features. FP3D is easy for extension because of the symbolic method used. It supports arbitrary equations, general coordinates and boundary conditions. It is a powerful test particle code for both 3D tokamaks and stellarators.

\section*{Acknowledgement}
We thank Dr S. Hirshman for the use of the 3-D equilibrium code VMEC code. We also thank Dr D. Gates and Dr C. Z. Zhu for use of the stellarator optimization code STELLOPT, and Dr M. Landreman for use of the drift-kinetic code SFINCS. This work is supported by the National MCF Energy R\&D Program of China (No. 2019YFE03050001).

\nocite{*}
\bibliography{ref}

\end{document}